\documentclass[usenatbib]{pasj01}\usepackage{natbib}

\newcommand{\bs}[1]{{\boldsymbol #1}}
\newcommand{\Mpch}{h^{-1}{\rm Mpc}}
\newcommand{\Msun}{h^{-1}{\rm M}_{\odot}}

\begin{document} 
\Received{}
\Accepted{}

\title{First results on the cluster galaxy population from the Subaru
  Hyper Suprime-Cam survey. II. Faint end color-magnitude diagrams and
  radial profiles of red and blue galaxies at $0.1<z<1.1$} 

\author{Atsushi J. \textsc{Nishizawa}\altaffilmark{1}%
}
\email{atsushi.nishizawa@iar.nagoya-u.ac.jp}
\author{Masamune \textsc{Oguri} \altaffilmark{2,3,4}}
\author{Taira \textsc{Oogi}\altaffilmark{4}}
\author{Surhud \textsc{More}\altaffilmark{4}}
\author{Takahiro \textsc{Nishimichi}\altaffilmark{4,5}}
\author{Masahiro \textsc{Nagashima}\altaffilmark{6}}
\author{Yen-Ting \textsc{Lin}\altaffilmark{7,8}}
\author{Rachel \textsc{Mandelbaum}\altaffilmark{9}}
\author{Masahiro \textsc{Takada}\altaffilmark{4}}
\author{Neta \textsc{Bahcall}\altaffilmark{10}}
\author{Jean \textsc{Coupon}\altaffilmark{11}}
\author{Song \textsc{Huang}\altaffilmark{4}}
\author{Hung-Yu \textsc{Jian}\altaffilmark{a}}
\author{Yutaka \textsc{Komiyama}\altaffilmark{12,13}}
\author{Alexie \textsc{Leauthaud}\altaffilmark{4,14}}
\author{Lihwai \textsc{Lin}\altaffilmark{7}}
\author{Hironao \textsc{Miyatake}\altaffilmark{15,4}}
\author{Satoshi \textsc{Miyazaki}\altaffilmark{12,13}}
\author{Masayuki \textsc{Tanaka}\altaffilmark{11}}
\altaffiltext{1}{Institute for Advanced Research, Nagoya University, Nagoya 464-8602, Aichi, Japan}
\altaffiltext{2}{Research Center for the Early Universe, University of Tokyo, Tokyo 113-0033, Japan}
\altaffiltext{3}{Department of Physics, University of Tokyo, Tokyo 113-0033, Japan}
\altaffiltext{4}{Kavli Institute for the Physics and Mathematics of the Universe (Kavli IPMU, WPI), University of Tokyo, Chiba 277-8582, Japan}
\altaffiltext{5}{CREST, JST, 4-1-8 Honcho, Kawaguchi, Saitama, 332-0012, Japan}
\altaffiltext{6}{Faculty of Education, Bunkyo University, 3337 Minami-Ogishima, Koshigaya-shi, Saitama 343-8511, Japan}
\altaffiltext{7}{Institute of Astronomy and Astrophysics, Academia Sinica, P.O. Box 23-141, Taipei 10617, Taiwan}
\altaffiltext{8}{Department of Physics, National Taiwan University, 10617 Taipei, Taiwan}
\altaffiltext{9}{McWilliams Center for Cosmology, Department of Physics, Carnegie Mellon University, Pittsburgh, PA 15213, USA}
\altaffiltext{10}{Department of Astrophysical Sciences, Princeton University, Princeton, NJ 08544, USA}
\altaffiltext{11}{Department of Astronomy, University of Geneva, ch. dE ́ cogia 16, 1290 Versoix, Switzerland}
\altaffiltext{12}{National Astronomical Observatory of Japan, 2-21-1 Osawa, Mitaka, Tokyo 181-8588, Japan}
\altaffiltext{13}{SOKENDAI(The Graduate University for Advanced Studies), Mitaka, Tokyo, 181-8588, Japan}
\altaffiltext{14}{Department of Astronomy and Astrophysics, University of California Santa Cruz, Santa Cruz, CA 95064, USA}
\altaffiltext{15}{Jet Propulsion Laboratory, California Institute of Technology, Pasadena, CA 91109, USA}



\maketitle

\begin{abstract}
  We present a statistical study of the redshift evolution of the
  cluster galaxy population over a wide redshift range from 0.1
  to 1.1, using $\sim 1900$ optically-selected CAMIRA clusters from
  $\sim 232$~deg$^2$ of the Hyper Suprime-Cam (HSC) Wide S16A data. 
  Our stacking technique with a statistical background subtraction
  reveals color-magnitude diagrams of red-sequence and blue cluster
  galaxies down to faint magnitudes of $m_z\sim 24$. We find that
  the linear relation of red-sequence galaxies in the color-magnitude
  diagram extends down to the faintest magnitudes we explore with a
  small intrinsic scatter $\sigma_{\rm int}(g-r)<0.1$. The scatter
  does not evolve significantly with redshift. The stacked
  color-magnitude diagrams are used to define red and blue galaxies in
  clusters for studying their radial number density profiles without
  resorting to photometric redshifts of individual galaxies. We find
  that red galaxies are significantly more concentrated toward cluster
  centers and blue galaxies dominate the outskirt of clusters.  We
  explore the fraction of red galaxies in clusters as a function of
  redshift, and find that the red fraction decreases with increasing
  distances from cluster centers. The red fraction exhibits a moderate
  decrease with increasing redshift. The radial number density
  profiles of cluster member galaxies are also used to infer the location of
  the steepest slope in the three dimensional galaxy density profiles.
  For a fixed threshold in richness, we find little redshift evolution
  in this location.
\end{abstract}

%
\section{Introduction}
\label{sec:introduction}
%

Clusters of galaxies are the largest gravitationally bound objects in
the Universe. Because their dynamics are determined mostly by gravity
and the spatial distribution of clusters also follows the large-scale
structure, clusters of galaxies are considered a good probe of 
cosmological structure. Therefore, understanding the formation history
of clusters is key for both studying the structure formation as well
as cosmological applications of clusters
\citep[e.g.][]{Rosati+:2002,Planelles+:2015,Castorina+:2014}.

One of the most prominent features of clusters of galaxies is the
presence of large number of red galaxies, which exhibit a
tight relation in the color-magnitude diagram (CMD)
\citep[e.g.][]{Stanford+:1998}.
This tight relation in the CMD, often referred as
the red-sequence, has been observed out to relatively high redshift,
$z\simeq 2$ \citep{Tanaka+:2010, Andreon+:2014, Cerulo+:2016, Romeo+:2016}.
The tilt of the color in the red-sequence can be explained mainly by
the decrease of metallicity for low mass galaxies
\citep{Faber:1973, KodamaArimoto:1997, DeLucia+:2007}, and
the age of the galaxies \citep{Ferreras+:1999, Gallazzi+:2006}.

Another notable property of clusters is that they also contain blue
galaxies. Thus the color distribution of cluster member galaxies has a
bimodal distribution \citep{Gilbank+:2007, Loh+:2008, Li+:2012}.
Historically, it is well known that there are more blue galaxies
observed at high redshifts than at low redshifts compared to the red
galaxies \citep{ButcherOemler:1978}. Also it has been found that
the abundance of morphologically different types of galaxies in clusters 
correlates with local density 
\citep{Dressler:1980,  Whitmore+:1991} and distance from the cluster
center \citep{Whitmore+:1993,George.etal:2013}.
These relations seen in the local universe can also be seen
in $z\sim 1$ clusters \citep{Postman+:2005}.
\citet{DePropris+:2004} showed that the fraction of blue galaxies in
blank field is a strong function of the local density with a sharp
decline near the density of the cluster environment toward $\sim 10\%$. 
This also suggests that member galaxies of clusters are dominated by
the red population, although there are still a non negligible amount of blue
galaxies in clusters. In \citet{Hennig+:2017}, Sunyaev-Zel'dovich (SZ)
detected clusters are stacked to derive the number density profile for
red and blue galaxies to find a clear difference in the concentration
between the two populations.

Because clusters of galaxies are rare objects, it has been difficult
to construct a large sample of clusters at high redshifts, which
requires both wide area and sufficient survey depth. Therefore
previous galaxy population studies of massive high-redshift
($z\sim 1$) clusters have been made using only a handful of clusters,
either from deep optical/infrared imaging surveys over small areas or
from follow-up imaging observations of X-ray and SZ-selected
high-redshift clusters.  For instance, IRAC Shallow Cluster Survey
(ISCS) includes 13 clusters at $1.01<z<1.49$ \citep{Snyder+:2012},
Gemini Cluster Astrophysics Spectroscopic Survey (GCLASS) has 10
spectroscopically confirmed rich clusters at $0.85<z<1.34$
\citep{Muzzin+:2012}, and HAWK-I Cluster Survey (HCS) contains only 9
clusters at $0.84<z<1.46$ \citep{Lidman+:2013}.
Conversely, there are large cluster samples at lower redshifts for
different data and different cluster finding methods.
\cite{GladdersYee:2000, GladdersYee:2005} uses two optical and near IR
filters to define the red sequence on Red-Sequence Cluster Survey
(RCS), 
\cite{Koester+:2007} applies red-sequence based finding method 
called maxBCG to SDSS galaxies, and
\cite{Rykoff+:2014} develops red-sequence based method with more
sophisticated algorithm and apply it on SDSS galaxies.

In this paper, we explore the average picture of the cluster galaxy
population using a large, homogeneous sample of clusters with a uniform
selection over a wide redshift range of $0.1 < z < 1.1$, which is
selected from the Subaru Hyper Suprime-Cam (HSC) survey. The sample
contains more than 1900 clusters optically identified by the
CAMIRA \citep[Cluster finding Algorithm 
based on Multi-band Identification of Red-sequence
gAlaxies;][]{Oguri+:2017} algorithm applied to the S16A internal data
release of the HSC survey. The unique combination of the area and
depth of the HSC survey is crucial for accurate galaxy population
studies for clusters at $z\sim 1$, as well as galaxy population
studies down to very faint magnitudes for low-redshift clusters. An
advantage of using a large homogeneous sample of clusters is that it
enables a statistical subtraction of projected foreground and
background galaxies, which is important for mitigating projection
effects in cluster galaxy studies.
With this unprecedented data set, we first examine the width of the
red-sequence of galaxies down to $m_z=24$ magnitude. Then we trace the 
redshift evolution of the radial profiles of red and blue galaxies in
clusters. It enables us to understand the formation history
and dynamical evolution of the clusters from $z=1.1$ to present.
We refer the interested reader to a series of complementary papers
describing different aspects of the properties of these cluster galaxies
\citep{Jian+:2017, Lin+:2017}.

This paper is organized as follows. In Section~\ref{sec:data}, we
describe the HSC photometric data and briefly overview the CAMIRA
cluster catalog. Our selection criteria for the photometric sample is
also presented.
In Section~\ref{sec:method}, we describe our method to identify
cluster member galaxies without using photometric
redshifts. Section~\ref{sec:redseq} presents the intrinsic scatter of
red galaxies down to the magnitude limit. The radial profile and
fraction of red member galaxies in clusters and their evolution over
redshifts are discussed in Section~\ref{sec:fred}. The comparison with
semi-analytical model results is also presented there. We summarize
our results in Section~\ref{sec:summary}. Unless otherwise stated, we
assume cosmological parameters as $\Omega_m=0.31, \Omega_\Lambda=0.69$
and $h=0.7$ that are consistent with
\citet{2014A&A...571A..16P} results throughout the paper.

%
\section{HSC data}
\label{sec:data}
%

\subsection{HSC photometric sample}
\label{ssec:hsc_photo}
The HSC survey started observing in March 2014 and is continuously
collecting photometric data over a wide area under
good photometric conditions \citep{HSCDR1:2017}.
The HSC survey consists of three different layers: Wide, Deep and
UltraDeep. In this paper, we use the photometric data from the Wide
layers of the internal S16A data release, which covers more than
200~deg$^2$ of the sky with five broadband filters ($grizy$).

In the S16A data, the Wide layer reaches the limiting magnitude of
26.4 in $i$-band. During the survey, $i$- and $r$-band filters were
replaced by newer ones, which have better uniformity over the entire
field of view \citep{Miyazaki+:2017, Kawanomoto+:2017}
 They are denoted $i2$ and $r2$ filters, respectively.
In this paper we do not discriminate the old vs new filters and simply denote
them as $i$ and $r$. For each pointing, we divide the exposure into 4
for $g$- and $r$-bands, and 6 for $i$-, $z$-, and $y$-bands with a
large dithering step of $\sim 0.6$~deg, to recover images at gaps
between CCDs and to obtain a uniform depth over the entire field.
The total exposure times for each pointing are $150\times 4=600$~sec
for $g$- and $r$-bands, and $200\times6=1200$~sec for $i$-, $z$-, and
$y$-bands. The expected 5$\sigma$ limiting magnitudes for the
$2''$ diameter aperture are 26.5, 26.1, 25.9, 25.1, and
24.4 for $g$-, $r$-, $i$-, $z$-, and $y$-bands, respectively. In Point
Spread Function (PSF) magnitudes, which is an aperture magnitude 
convolved with the measured and modeled PSF function,
we reach down to $i=26.4$ with
5$\sigma$ level. The median seeing of $i$-band images is $0\farcs61$
\citep{HSCOverview:2017}.

The images are reduced by the processing pipeline called \texttt{hscpipe}
 \citep{Bosch+:2017}, which is developed as a part of the
LSST (Large Synoptic Survey Telescope) pipeline 
\citep{Ivezic+:2008, Axelrod+:2010, Juric+:2015}.
The photometry and astrometry are calibrated in comparison with the
Pan-STARRS1 3$\pi$ catalog 
\citep{Schlafly+:2012, Tonry+:2012, Magnier+:2013}, which fully covers
the HSC survey footprint and has a set of similar filter response
functions. 
The photometry of the current version of hscpipe in crowded
  regions such as cluster centers is not accurate due to the
  complexity of object separation on the image, 
  \citep[\textit{deblending},][]{Bosch+:2017}.
For this reason, we combine two different methods
of measuring the photometry as described in detail in
Section~\ref{ssec:sample}.

\subsection{CAMIRA clusters}
\label{ssec:camira}
In this Section, we briefly overview the CAMIRA cluster finding
algorithm and properties of the catalog \citep{Oguri+:2017}.  CAMIRA
is a cluster finding algorithm based on the red-sequence galaxies
\citep{Oguri:2014}. By applying the algorithm to the HSC S16A data,
\citet{Oguri+:2017} constructed a catalog of $\sim 1900$ clusters from
$\sim 230$~deg$^2$ of the sky over the wide range of redshift
$0.1<z<1.1$ with almost uniform completeness and
purity. \citet{Oguri+:2017} first applied specific 
color cuts to a spectroscopic redshift-matched catalog to remove the
obvious blue  galaxies. Those galaxies are used only for calibrating
the color of the red-sequence and this color cut is not applied in cluster
finding itself. CAMIRA computes the likelihood of a galaxy being on the
red-sequence as a function of redshift, using a stellar population
synthesis (SPS) model of \citet{BC03}  with accurate calibration of colors
using the spectroscopic galaxies mentioned above. 
In contrast with the complex color degeneracy among different
parameters of SPS, CAMIRA simplify the parametrization to quantify the
red sequence galaxies.
They apply a single instantaneous burst at the formation redshift
$z_f=3$ and assume no dust attenuation, since these prescription is
good enough to represent the red-sequence color. The metallicity of
galaxies are modeled as logarithmic linear function of total
mass at $z=z_f$; i.e.
$\log Z_{\rm SPS}=\log Z_{11} + a_Z\log(M_{*}(z_f)/10^{11}{\rm M}_{\odot})$, with
$\log Z_{11}=-2, a_Z=0.15$. They additionally introduce the scatter of
metallicity to model the intrinsic scatter of the red-sequence with 
$\sigma_{\log Z}=0.14$ \citep{Oguri:2014}.
For individual galaxy, the likelihood is calculated by finding the
best fit parameters but color calibration is simultaneously applied by
using the spectroscopic sample, which recover the imperfect prediction
of SPS model \citep[see equation (2) of ][]{Oguri:2014}.
These likelihoods are
used to compute the richness parameter with which cluster candidates
are identified by searching for peaks of the richness. For each cluster
candidate, the BCG is identified by searching a bright galaxy near the
richness peak. The photometric redshift and richness are iteratively
updated until the result converges. As a result, the accuracy of the
photometric redshift reaches 1\% out to $z\sim 1$, which is better
than other cluster finding algorithms for different data set
\citep[e.g.][]{Rykoff+:2016,Mints.etal:2017}.

In this paper, we use 1902 CAMIRA clusters selected from $\sim
230$~deg$^2$ HSC Wide fields with richness $\hat{N}_{\rm mem}>15$. We
note that the number of CAMIRA clusters is slightly smaller than in
\citet{Oguri+:2017}, because we further impose the condition that all
galaxies have well measured $y$-band photometry. The difference is
less than 1\% and this does not change our results. Although there is
a large scatter in the relation between richness and mass,
\citet{Oguri+:2017} argued that the richness limit of $\hat{N}_{\rm
  mem}=15$ roughly corresponds to a constant mass limit of
$M_{200m}=10^{14} \Msun$. Due to the large scatter of the
mass-richness relation and lack of the mass calibration from weak
lensing, in this paper we do not divide the sample in richness bins.
In contrast, in order to study the redshift evolution of the cluster
galaxy property, we divide these clusters into different redshift bins
with a bin size of $\Delta z=0.05$ at $z<0.4$ and $\Delta z=0.1$ at
$z\geq 0.4$. Table~\ref{tab:camira_num} summarize the mean redshift of
clusters and the number of clusters with richness 
$\hat{N}_{\rm mem}>15$ in each redshift bin.  
The effect of cluster evolution on sample selection will be
  addressed in a future paper once the halo mass estimates from weak
  lensing will be available. The different sample selection can be found 
  in our series of paper \citep{Lin+:2017}.

\begin{figure}[th]
  \begin{center}
  \includegraphics[width=\linewidth]{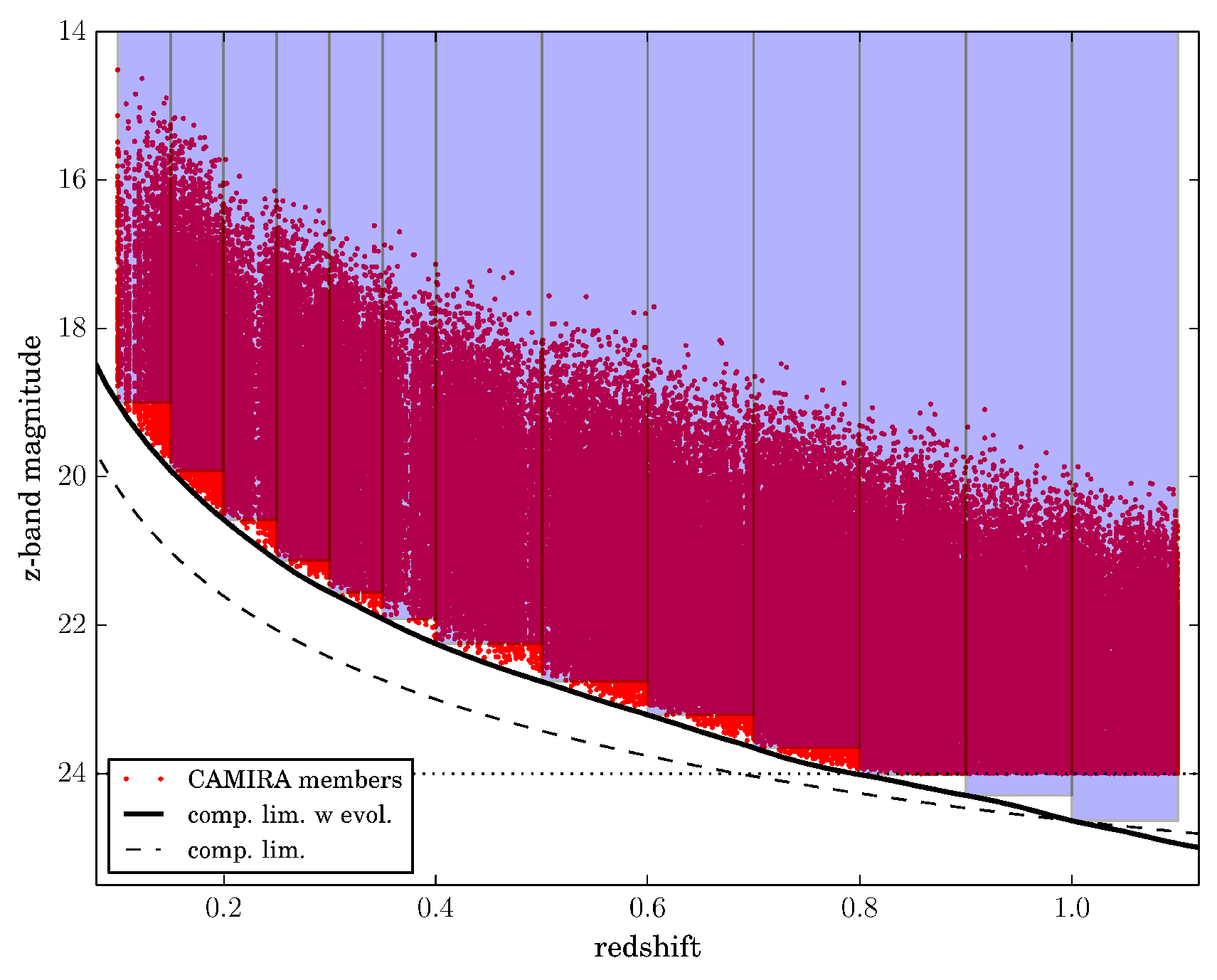}
  \caption{
    Redshift and magnitude distribution of all CAMIRA member galaxies.
      The dashed line represents the observer-frame $z$-band magnitude 
      of a SPS with constant absolute magnitude $M_z$=-18.5 as seen at 
      different redshifts. The thick solid line is the same constant 
      absolute magnitude but after applying $K$-correction and taking 
      into account passive evolution.
    Shaded rectangles are regions
    where the galaxy sample is complete and suitable for exploring the redshift
    evolution of clusters in various aspects over different redshifts.
    The horizontal dotted line represents the $m_z=24.0$ mag cut
    that we apply to remove the galaxies affected by large
    photometric errors especially in the highest redshift bins of
    the cluster sample (see Section~\ref{ssec:sample} for details).
    \label{fig:camira}}
  \end{center}
\end{figure}

\begin{table}
  \begin{center}
    \begin{tabular}{cccc} \hline\hline
      redshift bin & mean redshift & color & number of clusters \\ \hline
      0.10--0.15  & 0.13 & $g-r$ & 43  \\
      0.15--0.20  & 0.18 & $g-r$ & 91  \\
      0.20--0.25  & 0.22 & $g-r$ & 74  \\
      0.25--0.30  & 0.28 & $g-r$ & 145 \\
      0.30--0.35  & 0.32 & $g-r$ & 130 \\
      0.35--0.40  & 0.38 & $g-r$ & 70  \\
      0.40--0.50  & 0.45 & $r-i$ & 192 \\
      0.50--0.60  & 0.55 & $r-i$ & 220 \\
      0.60--0.70  & 0.65 & $r-i$ & 179 \\
      0.70--0.80  & 0.75 & $i-y$ & 231 \\
      0.80--0.90  & 0.85 & $i-y$ & 217 \\
      0.90--1.00  & 0.95 & $i-y$ & 137 \\
      1.00--1.10  & 1.05 & $i-y$ & 173 \\ \hline\hline
    \end{tabular}
  \end{center}
  \caption{
      The definition of redshift bins used in this paper. For each
      redshift bin, the mean redshift of clusters, the color 
      combination used to define red and blue galaxies, and number of
      clusters with $\hat{N}_{\rm mem}>15$ are shown.
      \label{tab:camira_num}}
\end{table}

\subsection{HSC sample selection}
\label{ssec:sample}
In this Section, we describe our selection of the HSC photometric galaxy
sample, which is used for our analysis of the cluster galaxy
population. In this paper, we use magnitudes for each object that are
derived by combining CModel magnitudes \citep{Abazajian+:2004, Bosch+:2017}
with PSF matched aperture magnitudes, the so-called afterburner
photometry \citep{Bosch+:2017}. The CModel magnitude is
obtained by fitting the object light profile with the sum of 
a de Vaucouleurs bulge and an exponential disk convolved with
the PSF. PSFs are measured at the positions of stars and then modeled
to interpolate over the entire field of view \citep{Bosch+:2017}.  
The PSF-matched aperture magnitude in the afterburner photometry is
obtained by stacking the image after blurring each exposure toward the
target PSF size and measuring the photometry at a given aperture size. 
All the magnitudes are corrected for the Galactic extinction \citep{SFD:1998}.
First we define the total magnitude 
\footnote{Note that the CModel magnitude is not a total magnitude when
the deblender fails, which may often occur in crowded regions like cluster centers.}
of each galaxy with the $z$-band
CModel magnitude. Then we derive the magnitudes in other bands
as
\begin{equation}
  m_{x} = 
  m_{z}^{\rm CM}
  + 
  \left(
    {m}_{x}^{\rm ab} - {m}_{z}^{\rm ab}
  \right),
\end{equation}
where ${m}_{z}^{\rm CM}$ is the CModel magnitude in the $z$-band measured
with forced photometry on the PSF-unmatched coadd image
(\texttt{cmodel\_mag}), and ${m}_{x}^{\rm ab}$ (${m}_{z}^{\rm ab}$) is
the PSF-matched aperture magnitude in $x$-band ($z$-band) measured in
$0\farcs55$ aperture in radius on the stacked image, where the PSF is
convolved to homogenize the target PSF size of $1\farcs1$
(\texttt{parent\_mag\_convolved\_2\_0}).
As the error of the afterburner photometry is significantly 
underestimated because the neighboring pixels are highly correlated
due to the blurring, we use the photometric error associated with
the PSF-unmatched aperture photometry with a corresponding aperture 
instead of the afterburner photometry error.

\begin{figure}[th]
  \begin{center}
  \begin{tabular}{c}
    \includegraphics[width=\linewidth]{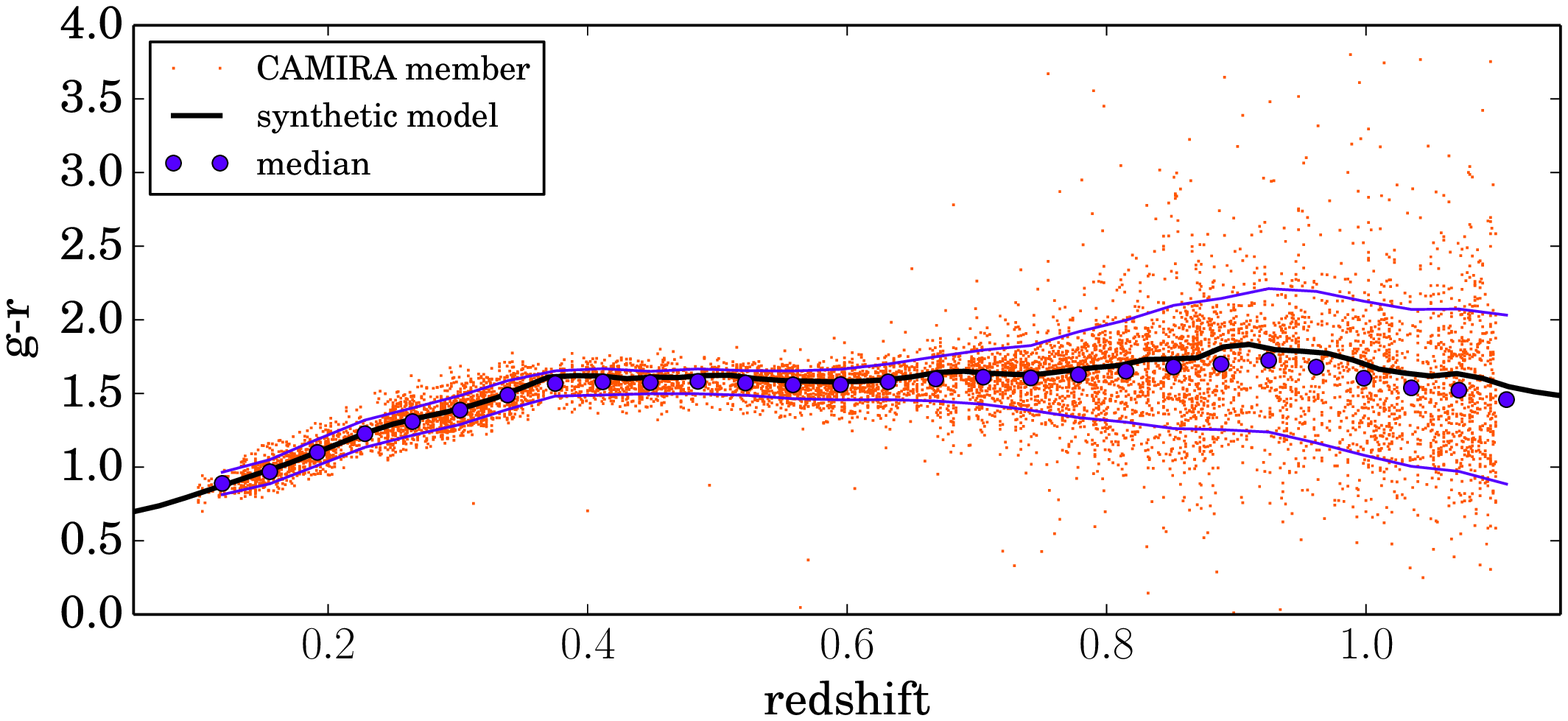}\\
    \includegraphics[width=\linewidth]{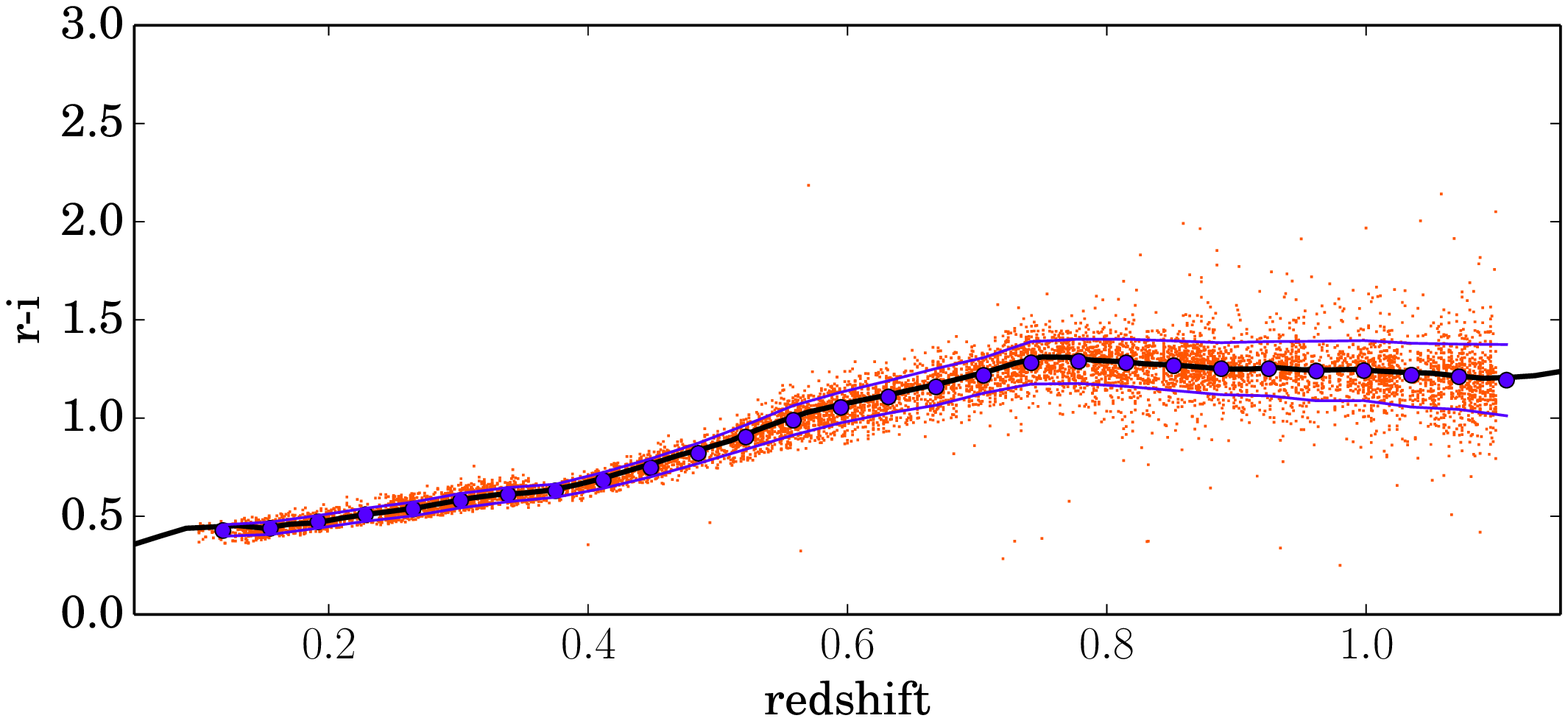}\\
    \includegraphics[width=\linewidth]{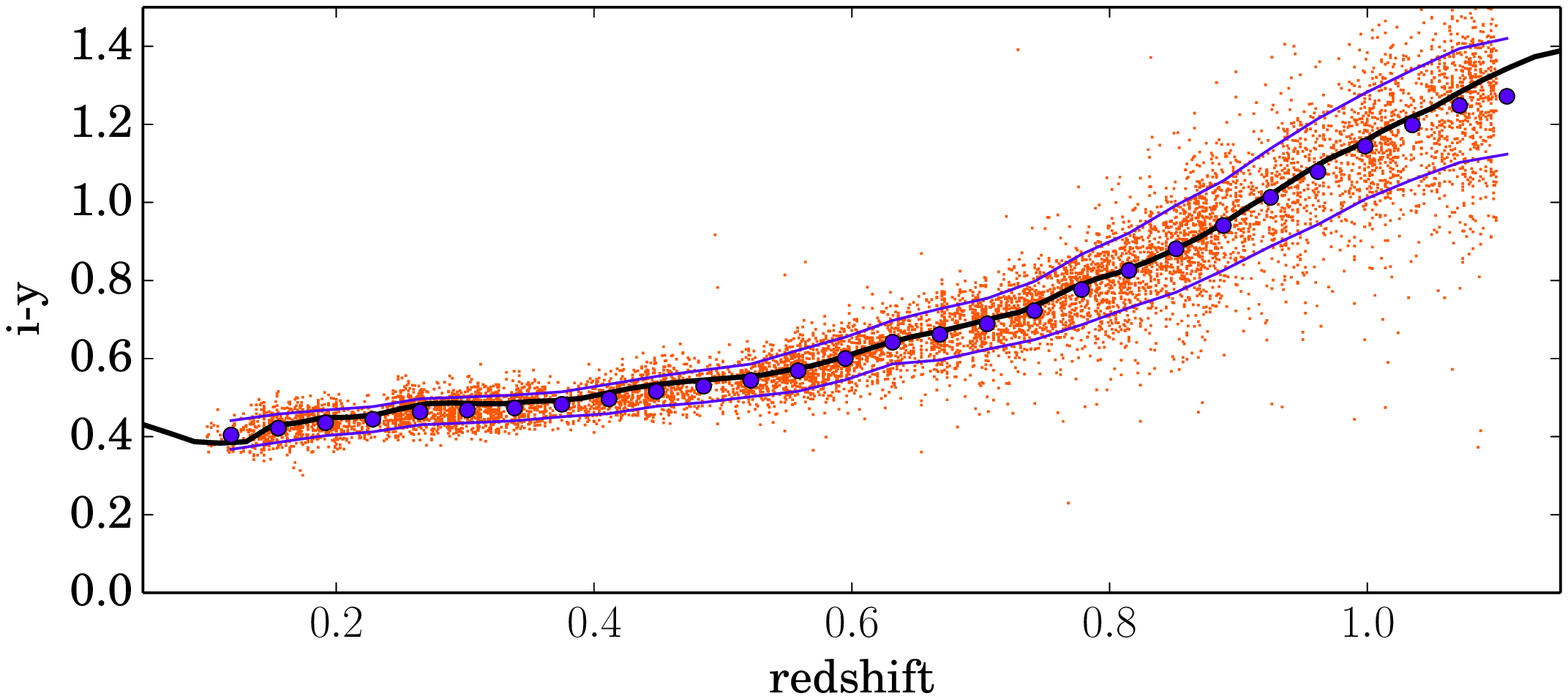}\\
  \end{tabular}
  \caption{Redshift-color relation of red-sequence galaxies.
    Red points are cluster member galaxies identified by CAMIRA, and
    their median and 1$\sigma$ region are denoted by filled circles
    and thin solid lines. Thick solid line show model colors from the
    SPS model \citep{BC03} calibrated with spectroscopic redshifts.
    \label{fig:rs_redshift}}
\end{center}
\end{figure}

In the following we describe flags applied to select galaxies with
high-quality photometry \citep{HSCDR1:2017}.\\ 
\textbf{Flags in forced photometry table}
\begin{itemize}
  \item{} \texttt{[grizy]flags\_pixel\_edge} is not True 
  \item{} \texttt{[grizy]flags\_pixel\_interpolated\_center} is not True 
  \item{} \texttt{[grizy]flags\_pixel\_cr\_center} is not True\\
    We apply the above three constraints to avoid objects geometrically
    overlapping with the masked region [\texttt{EDGE} or \texttt{NO
      DATA}], or object centers close to interpolated pixels or
    suspected cosmic rays.
  \item{} \texttt{[grizy]cmodel\_flux\_flags} is not True \\
    These flags ensure that the CModel flux is 
    successfully measured.
  \item{} \texttt{[grizy]centroid\_sdss\_flags} is not True \\
    Exclude objects for which measurements of centroids failed using
    the same method as in the Sloan Digital Sky Survey \citep{Bosch+:2017}.
  \item{} \texttt{[gr]countinputs} $>1$ 
  \item{} \texttt{[izy]countinputs} $>3$
    In the HSC-Wide, we divide every pointing into 4 for $g$- and
    $r$-band, and 6 for $i$-, $z$- and $y$-bands. Here we use objects
    taken twice or more for $g$ and $r$, and four times or more for
    $i,z$ and $y$ bands. 
  \item{} \texttt{detect\_is\_primary} is True \\
    We also remove blended objects to avoid
    ambiguous photometric measurements.
  \item{} \texttt{zcmodel\_mag} - \texttt{a\_z} $<24.0$ \\
    We limit our sample in these magnitudes ranges so that all the
    objects have high signal to noise ratio. 
  \item{} \texttt{rcmodel\_mag} - \texttt{a\_r} $<28.0$
  \item{} \texttt{icmodel\_mag} - \texttt{a\_i} $<28.0$
    These two flags are not for the primary object selection but 
    for removing too faint objects.
  \item{} \texttt{zcmodel\_mag\_err} $<0.1$ \\
    We also directly impose the $S/N$ cut corresponding to 
    $S/N \gtrsim 10$.
  \item{} \texttt{iclassification\_extendedness} $=1$\\
    Stars are excluded.
\end{itemize}
\textbf{Flags in afterburner table}
\begin{itemize}
  \item{} \texttt{[grizy]parent\_flux\_convolved\_2\_0\_flags} is not True 
\end{itemize}
In addition to the above selection, we use only galaxies brighter than
the limits shown in Figure~\ref{fig:camira} depending on the redshift
of clusters.

As a sanity check, we compare colors of CAMIRA cluster member galaxies
with model colors from the SPS model used in CAMIRA cluster finding
\citep{Oguri+:2017}, which is based on the SPS model of \citet{BC03}
with the calibration of colors from spectroscopic redshifts as
described above. Figure~\ref{fig:rs_redshift} shows the redshift
evolution of colors of CAMIRA cluster member galaxies, where we derive
colors of cluster member galaxies by matching the photometric galaxy
sample constructed above with a catalog of CAMIRA member galaxies from
\citet{Oguri+:2017}. We find that model colors and median colors of
the member galaxies agree well, as expected. As shown in
Figure~\ref{fig:rs_redshift}, in the redshift ranges of $0.1<z<0.4$,
$0.4<z<0.7$, $0.7<z<1.1$, $g-r$, $r-i$, and $i-y$ respectively show
rapid color changes, because these 
colors cover the $4000\AA$ break at these redshifts. Moreover, we find
that the above combination of filters shows the tightest scatter around
the theoretical prediction in each redshift range. Therefore, we
use these colors to construct the color magnitude diagram to see the
red-sequence galaxies at fainter magnitude with a stacking analysis (see
Table~\ref{tab:camira_num}).

%
\section{Statistical identification of cluster member galaxies}
\label{sec:method}
%
\subsection{Stacking analysis}
\label{ssec:stacking}
In this paper, we study distributions of cluster member galaxies
statistically, without resorting to spectroscopic or photometric
redshifts. 
Since we do not use both photometric redshift and spectroscopic
redshift for each galaxy, physical quantities of individual galaxies
such as stellar masses or star formation rates are not available;
however, an advantage of our approach is that we can
exclude any uncertainties associated with photometric redshift
measurements, which may also induce uncertainties in identification
of cluster member galaxies. This statistical method has been used 
previously in the literature \cite[e.g.][]{Lin+:2004, Hansen+:2005,
  LohStrauss:2006}. Here we describe our specific procedure.

First we divide the CAMIRA cluster catalog into subsamples at
different redshift bins, as shown in Table~\ref{tab:camira_num}.
In each redshift bin, we have roughly $\sim 50-200$ clusters. The
galaxy distribution associated with the $i$-th cluster and $j$-th
annulus can be written as
\begin{equation}
  \label{eq:gal_count}
  N_{ij}^{\rm in}(z_{\rm cl})
    =\sum_k \Theta(|\Delta \bs{\theta}_{ik}|\chi_{{\rm cl},i}-r_{P,j})
    \Theta(r_{P,j+1}-|\Delta \bs{\theta}_{ik}|\chi_{{\rm cl},i}),
\end{equation}
where the summation $k$ runs over galaxies, $\chi_{\rm cl}$ is
comoving distance to the cluster redshift $z_{\rm cl}$, 
$\Delta \bs{\theta}_{ik}$ is the sky separation of the $k$-th galaxy
with respect to the $i$-th cluster center, and $\Theta$ is a Heaviside
step function. We consider the comoving radial distance from the cluster
center $r_P$ in the range from 0.1 to 5~$\Mpch$ (comoving), which is
divided into 15 logarithmically uniform bins. After stacking over all
clusters within each redshift bin, we have
\begin{equation}
  \label{eq:gal_in}
  N_j^{\rm in}(r_P)
  = 
  \sum_{i} N_{ij}^{\rm in}.
\end{equation}
This number $N_j^{\rm in}$ includes not only cluster member galaxies
but also foreground and background galaxies along the line of
sight. In order to remove these foreground and background galaxies, we 
assume that the galaxy number distribution outside the cluster region
defined by $r=|\Delta \bs{\theta}_{ik}|\chi_{{\rm cl},i}>5 \Mpch$ well
represents the distribution of the foreground and background galaxy
population. Although the distribution may differ field by field due to
both inhomogeneous observing conditions and large-scale structure of the
Universe, such local variation of the galaxy number distribution is
expected to be averaged out after stacking many clusters at
different positions on the sky, as long as the sky coverage of the
survey is sufficiently large \citep[e.g.,][]{Goto+:2003}.
The galaxy number outside the cluster region $N^{\rm out}$ for
each redshift bin is estimated by using all galaxies that are located
at $r>5 \Mpch$ for all the clusters in the redshift bin. 
The number of galaxies is rescaled by the area before subtraction. 
We then subtract the contamination by foreground and background
galaxies using $N^{\rm out}$ estimated above.
We measure the areas occupied by the galaxies outside the cluster regions by
counting the number of randoms in the random catalog
\citep{Coupon+:2017}.
The random catalog is created based on the pixel-based information
and inherits most of the photometric flags on the object images. One
can find the corresponding version of random catalog to the object
catalog in the same data release site.
The random catalog takes account of both the selection criteria described in
Section~\ref{ssec:sample} and masks. Now the foreground and background
subtracted number of galaxies can be written as,
\begin{equation}
  N_j = N^{\rm in}_j - N^{\rm out} \frac{R^{\rm in}_j}{R^{\rm out}},
\end{equation}
where $R$ is the number of randoms which is defined in the exactly
same manner as in equations~\ref{eq:gal_count} and \ref{eq:gal_in}.

\subsection{Color correction}
\label{ssec:color_corr}
As shown in Figure~\ref{fig:rs_redshift}, the tight relation of
red-sequence evolve with redshift. This means that galaxy colors
evolve with redshift, even within the same redshift bin.  In order to
obtain accurate stacking results, including accurate separation of
red and blue galaxies based on their colors, we apply a correction for the
color evolution as a function of redshift before stacking many 
clusters to study the population of red and blue galaxies within each
redshift bin.
For clusters at $z=z_1$, we derive the corrected color as
$C_{\rm corr}=C_{\rm obs}-C^{\rm th}(z_1)+C^{\rm th}_{\star}$,
where $C^{\rm th}$ denotes the theoretically derived galaxy color
based on the stellar population synthesis model of \citet{BC03}
with the calibration of colors using spectroscopic galaxies in the HSC
survey \citep{Oguri+:2017}. 
$C_{\star}^{\rm th}$ is $C^{\rm th}$ at the median redshift $z_{\star}$ within
the redshift bin.
Furthermore, we correct the color gradient
as a function of magnitude using $z$-band magnitudes.
We fit the colour-magnitude relation along the red sequence
by the linear function as 
$g(m_z)=g_{\star} + \alpha (m_z - m_{z,\star})$, and correct the colors of
all galaxies to the 
red-sequence zero-point (intercept of the linear relationship)
estimated at the median magnitude of the cluster member galaxies,
$g_{\star}\equiv g(m_{z, \star})$.

To summarize, the corrected color of a galaxy with magnitude $m_z$ for
a cluster at $z_{\rm cl}$ stands for the color difference from that of
the red sequence and
is derived from the observed raw color $C_{\rm obs}$ as
\begin{equation}
  \label{eq:color_corr}
  C(m_z | z_{\rm cl}, m_{z,\star})
  =
  C_{\rm obs} -C^{\rm th}(z_{\rm cl})+C^{\rm th}_{\star}
  - g(m_z) + g_{\star}.
\end{equation}
We show the color gradient in terms of redshift in
Figure~\ref{fig:rs_redshift} for CAMIRA member galaxies and the color
gradient in terms of z-band magnitude in
Figure~\ref{fig:cmd_bgsubtracted1}.

\subsection{Definition of Red and Blue galaxies}
\label{ssec:def_redblue}
\begin{figure*}[th]
  \begin{center}
    \includegraphics[width=\linewidth]{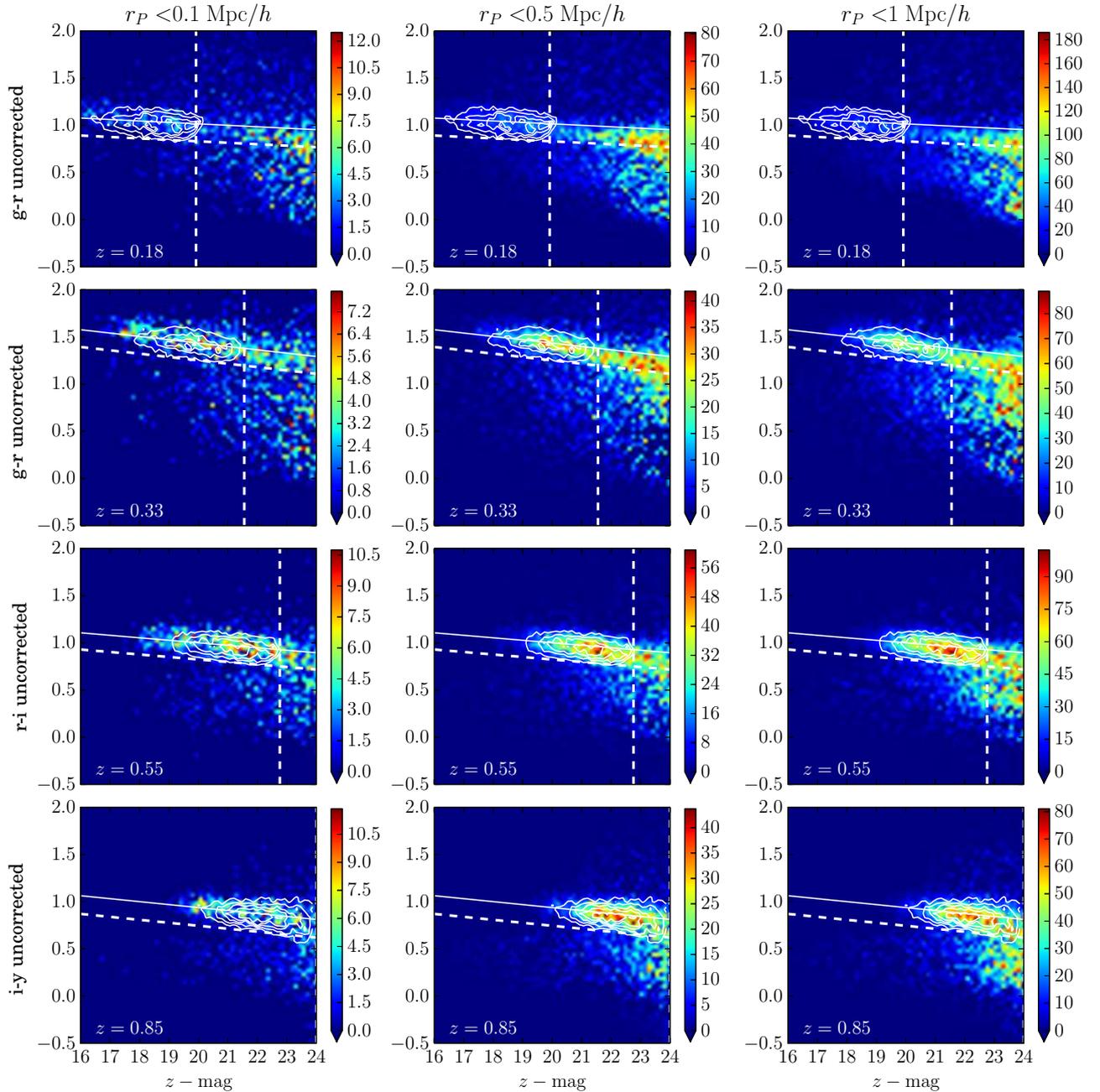}
  \end{center}
  \caption{Color-magnitude diagrams (CMDs) that are derived by stacking
    photometric galaxies over all CAMIRA clusters. Color level stands
    for the number of galaxies in each cell after foreground and
    background galaxies are statistically subtracted (see the text
    for details).
    From left to right, we show CMDs for the cluster centric radius
    $r_P<0.1, 0.5$ and $1.0 \Mpch$.
    From top to bottom, the mean redshift of
    the clusters are $z_{\rm cl}=0.18$, $0.33$, $0.55$, and $0.85$.
    Overlaid contours in each panel are the distribution of cluster
    member galaxies identified by CAMIRA. Solid lines show the slopes
    that minimize the scatter of CAMIRA member galaxies around the
    line, i.e. $g(m_z)$ correction. We define red and blue galaxies
    for each redshift bin by those above and below the dashed line
    (which is defined by a line 2$\sigma$ below the solid line),
    respectively. Vertical dotted lines are the apparent magnitude cut
    corresponding to the rest frame $M_z<-18.5$.
    \label{fig:cmd_bgsubtracted1}}
\end{figure*}

\begin{figure}
  \includegraphics[width=\linewidth]{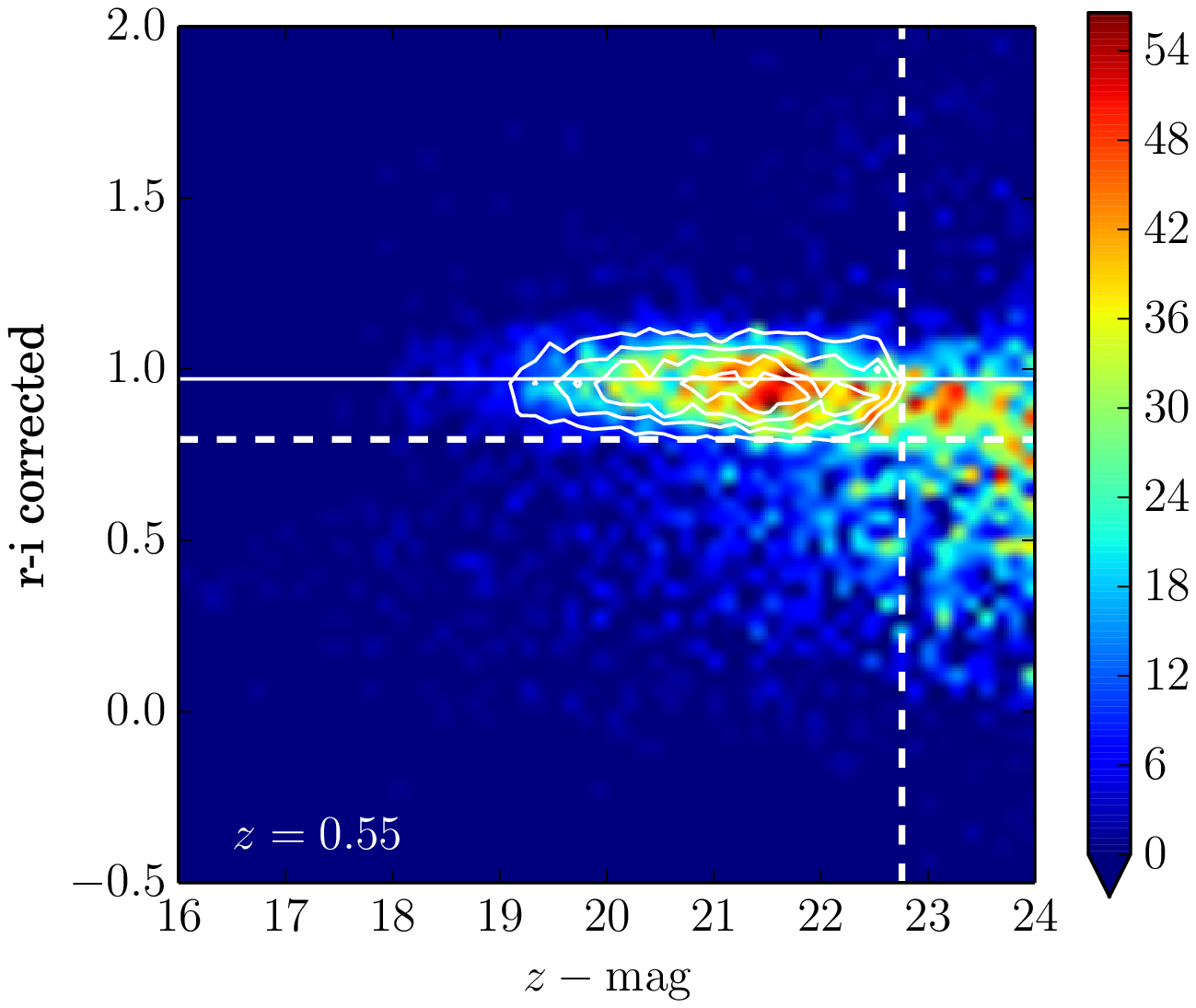}
  \caption{
    Same as Fig.~\ref{fig:cmd_bgsubtracted1} for $z=0.55$ and
    $r_P<0.5\Mpch$, but color is corrected according to equation
    \ref{eq:color_corr}.
  \label{fig:cmd_corr}}
\end{figure}

While there are a variety of definitions of red and blue galaxies in
the literature, we introduce an empirical definition based on the
observed data. Since CAMIRA cluster member galaxies represent the
population of quiescent galaxies, the location of the CAMIRA cluster
member galaxies in the CMD is well localized. This means, at a given
redshift, galaxies having different star formation activities have
different colors. We define blue galaxies as those that are located in
the CMD 2$\sigma$ away (on the bluer side) from the linear relation of
the red-sequence obtained in Section~\ref{ssec:color_corr}. 
Figure~\ref{fig:cmd_bgsubtracted1} shows CMD after the foreground and
background subtracted in different redshifts (low-z to high-z from top
to bottom) and different cluster centric radius (inner to outer from
left to right). The color tilts are not corrected (but see
Fig.~\ref{fig:cmd_corr} for color corrected diagram for $z=0.55$
and $r_P<0.5\Mpch$).
Horizontal dashed lines are the location dividing the
sample into red and blue galaxies.
It is clearly seen that there are few blue galaxies at inner region
of clusters and it increases with the cluster centric radius.
We will see this more in detail in Section~\ref{sec:fred}. We note
that if we carefully focus on the faint end of the CMD, the linear
function obtained by CAMIRA member galaxies are slightly off from the
peak of the red galaxies distribution. 
Unlike the CAMIRA member galaxies which are well matured to be red
sequence, faint galaxies near the red-sequence track still in the
stage of star forming and in the transition phase from star forming
galaxies to quiscent galaxies. For the thorough investigation, we need
to divide the cluster sample in finner mass bins, which will be
devoted to our future paper.

%
\section{Red-sequence at the Faint End}
\label{sec:redseq}
%
In this Section, we study the red-sequence galaxies within cluster
centric radius $r_P<0.5\Mpch$ at the very faint
end down to $m_z\sim 24$, which is enabled by our careful statistical
subtraction of foreground and background galaxies. Specifically, we
study how the scatter of the red-sequence changes as a function of
magnitude. We model the color-corrected, foreground and
background-subtracted CMD distribution with the following double
Gaussian \cite[e.g.][]{Hao+:2009}
\begin{eqnarray}
  \label{eq:dgauss}
  &&n(C|m_z)
  =
     \frac{A_R(m_z) }{\sqrt{2\pi\sigma_R^2(m_z)}}
     \exp
     \left[ -\frac{(C-C_R)^2}{2\sigma_R^2(m_z)} \right]
     +
     \nonumber \\
  &&\hspace{8em}
     \frac{A_B(m_z)}{\sqrt{2\pi\sigma_B^2(m_z)}}
     \exp
     \left[ -\frac{(C-C_B)^2}{2\sigma_B^2(m_z)} \right]
\end{eqnarray}
where 
the parameters
$A_x, \sigma_{x}$ and $C_x$ with $x$ being either $R$ (red) or $B$
(blue), are treated as free parameters. As we already corrected for
the color tilt against the magnitude in Section~\ref{ssec:color_corr},
the mean of the red component, $C_R$, can well be described by a
constant.
For simplicity, we also assume that the blue
component has constant mean ($C_B$), which is equivalent to assuming
that the tilt of the color-magnitude relation for blue galaxies is the
same as that for red galaxies. This assumption is reasonable because 
the blue galaxies do not have tight relation with the $m_z$ but are rather
broadly distributed and thus 
insensitive to the choice of color correction;
as far as the tilt correction is linear, the color correction simply
changes the $C_B$ and $\sigma_B$ at each $m_z$ bin 
but it does not
affect the estimate of $\sigma_R$ we are interested in.
We divide the CMD in several magnitude bins, and estimate the values
of $\sigma_R$ for each magnitude bin with Markov-Chain Monte-Carlo
method by keeping other parameters free but fixing the $C_R$ to its
corrected value obtained in Section~\ref{ssec:color_corr}.

\begin{figure}[th]
  \begin{center}
    \includegraphics[width=\linewidth]{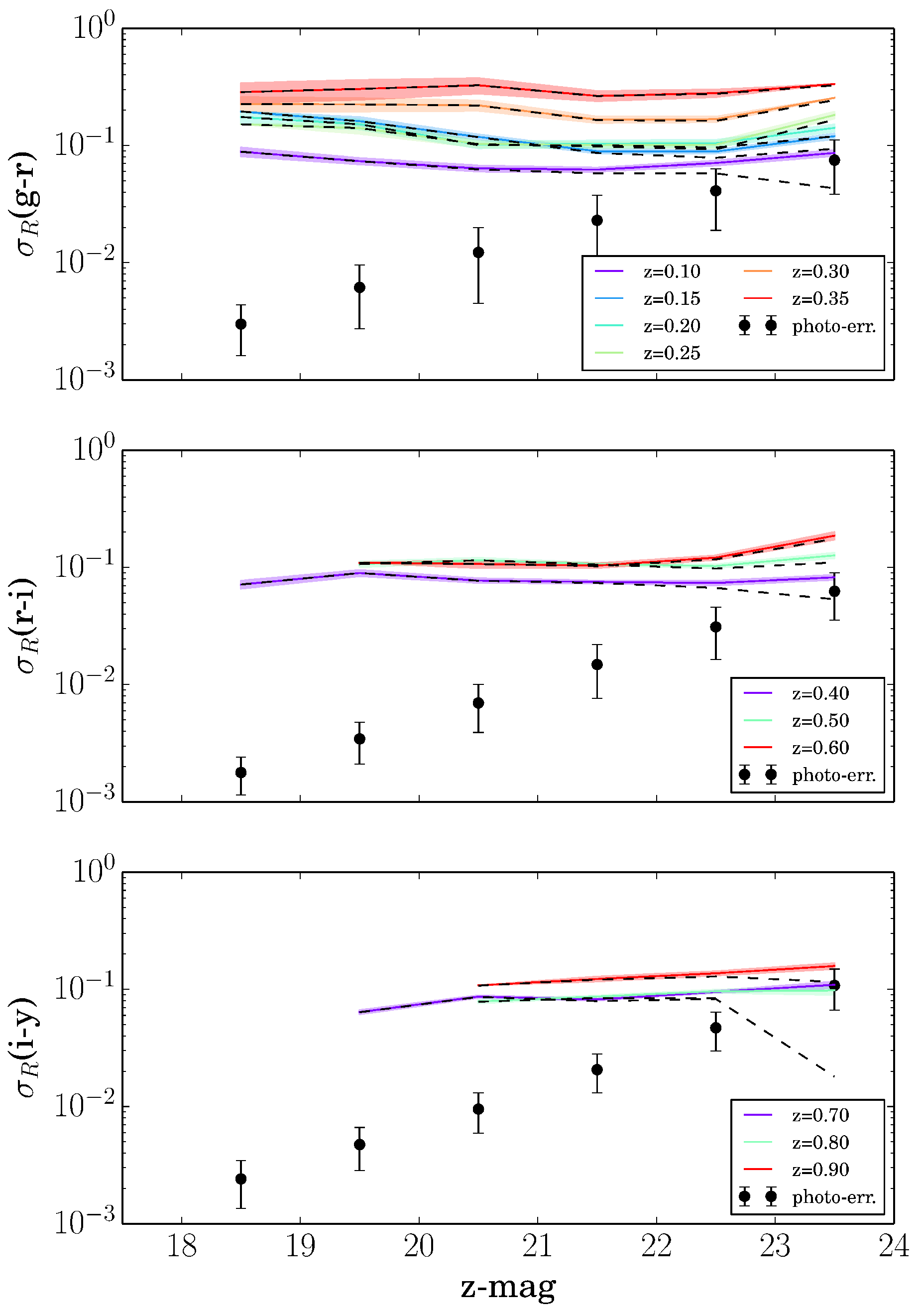}
  \end{center}
  \caption{The scatter of colors of red-sequence galaxies as a
    function of $z$-band magnitude. Results are shown for different
    redshift bins. Shaded regions show observed scatter, whereas
    dashed lines show the estimated intrinsic scatter after
    subtracting the scatter due  to the photometric error which is
    shown by filled circles with errorbars. The circles and errorbars
    are median and one sigma scatter of the photometric error in each
    magnitude bin.
    \label{fig:rs_scatter}}
\end{figure}

Figure~\ref{fig:rs_scatter} shows the best fit scatter parameter
$\sigma_R$ as a function of magnitude, for different cluster
redshifts. As discussed above, we use different colors for clusters at
different redshift, such that these colors refer to approximately the
same color in the cluster rest frame. At the very faint end, we need 
to take account of the scatter associated with the photometric error,
which has a significant contribution to the observed scatter at
$m_z\sim 24$. As the intrinsic scatter is not correlated with the
photometric error, we can separate their contributions as 
\begin{equation}
  \sigma_{\rm obs}^2
  =
  \sigma_{\rm photo}^2  + \sigma_{\rm int}^2,
\end{equation}
where $\sigma_{\rm obs}, \sigma_{\rm photo}$ and $\sigma_{\rm int}$
are the observed scatter, the scatter due to the photometric error,
and the intrinsic scatter of the red-sequence, respectively.
We find that the intrinsic scatter of the red-sequence galaxies after
subtracting the photometric error is almost constant over wide range
of magnitudes. We also find that there is no significant redshift
evolution of the scatter, 
which is consistent with the previos work using smaller sample of
clusters \citep[e.g.,][]{Cerulo+:2016, Hennig+:2017}. The Figure
suggests a slight decrease of the scatter at the faint end, but 
being the photometric error large at faint magnitudes, the
intrinsic scatter is likely to be underestimated.
Over most of the magnitude range, however,
the photometric error is much smaller than the intrinsic scatter,
which indicates that our result is robust against the photometric
error. 

%
\section{Cluster Profile and Fraction of Red galaxies}
\label{sec:fred}
%

\subsection{Cluster Profile}
\label{ssec:fred-r}
Given the timescale for the evolution of galaxies, tracking the redshift
evolution of the number density profiles for red and blue components can help
us understand the dynamical history of the formation of galaxy clusters.  The
radial mass density profile of dark matter halos has long been thought to have
a long tail that goes like $\rho\propto r^{-3}$ \cite{NFW}. With such
a profile, the total enclosed mass of a cluster diverges
logarithmically, and the total mass associated with the halo depends
upon the arbitrary boundary imposed on the halo.  The splashback
radius, marked by the apocenter of the recently infalling material,
provides a clear physical boundary for the halo and can be used to
identify the edges of dark matter halos \citep{DK:2014,More+:2015}.
The splashback radius manifests itself as a sharp drop in the matter
density at its location \citep{DK:2014, Adhikari:2014}.  The dark
matter profile with such a density jump can be modeled with an inner
universal profile multiplied by a transition function and an outer
profile which represents the so-called two-halo contribution
\citep{DK:2014}. This can be explicitly written as
\begin{eqnarray}
  \label{eq:fullprof1}
  &&\rho(r)
     =
     \rho^{\rm in}(r) \left[ 
     1+\left( \frac{r}{r_t} \right)^\beta 
     \right]^{-\gamma/\beta}
     +
     \rho^{\rm out}(r), \\
  \label{eq:fullprof2}
  &&\rho^{\rm out}(r)
     =
     \rho_m \left[
     b_e \left( \frac{r}{5R_{200}}\right)^{-s_e} + 1
     \right],
\end{eqnarray}
where $r_t, \gamma$ and $\beta$ denote the location of the dip in 
the profile, the steepness of the dip and how rapidly the slope changes, 
respectively. They all characterize the transition between inner and outer 
profiles. For the outer profile, $\rho_m, b_e$ and $s_e$ represent the overall 
normalization, relative normalization of power law profile and index of 
the power law, respectively.
\citet{DK:2014} express $r_t/R_{\rm 200m}$ as a function of the accretion rate
$\Gamma$ as $r_t/R_{\rm 200m} = (0.62+1.18\exp[-2\Gamma/3])$, but in this paper
we keep $r_t$ as a free parameter since we do not have reliable estimate of the
either the $R_{\rm 200m}$ or the accretion rate of our optically-selected
clusters. We use the NFW \citep{NFW} profile to describe the inner profile, 
\begin{equation}
  \label{eq:fullprof3}
  \rho^{\rm NFW}(r)
  =
  \frac{\rho_s}{(r/r_s) (1+r/r_s)^2},
\end{equation}
where $r_s$ and $\rho_s$ denote the transition scale of slope
  from $-1$ to $-3$ and overall normalization, respectively.
While we use $r_t$ for fitting observed radial profiles, following
\citet{More+:2015} we define a splashback radius, $R_{\rm sp}$, as the radius
where the radial profile attains its steepest slope. 
As in \citet{More+:2016}, we allow $\beta$ and $\gamma$ to take
the values with $\log \beta=\log 6 \pm 0.2$ and 
$\log \gamma = \log 4 \pm 0.2$.
The profiles of equations \ref{eq:fullprof1},
\ref{eq:fullprof2} and \ref{eq:fullprof3} are numerically integrated
along the line of sight to project to the two-dimensional sky.

\begin{figure*}[th]
  \begin{tabular}{cc}
    \includegraphics[width=0.5\linewidth]{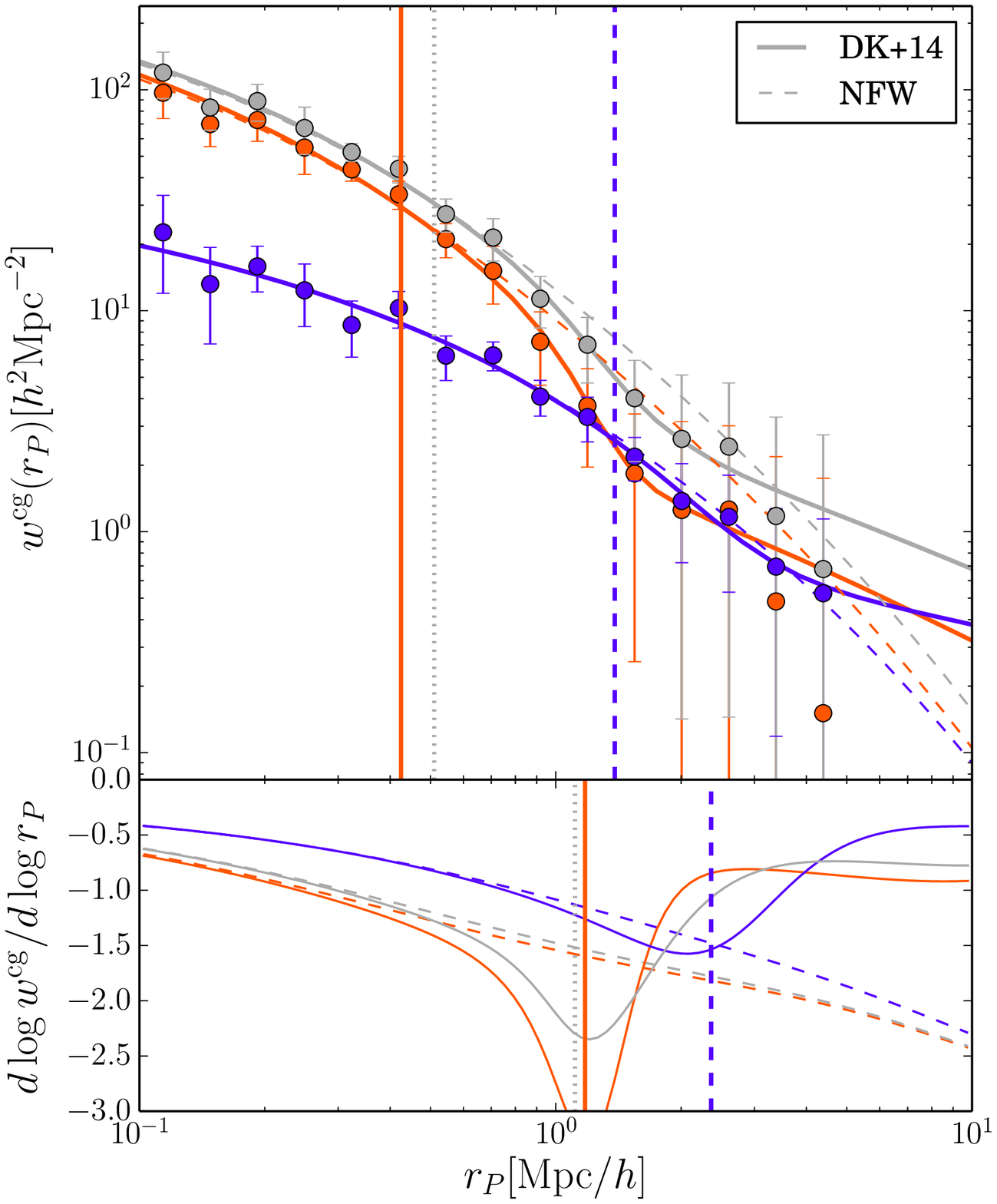}&
    \includegraphics[width=0.5\linewidth]{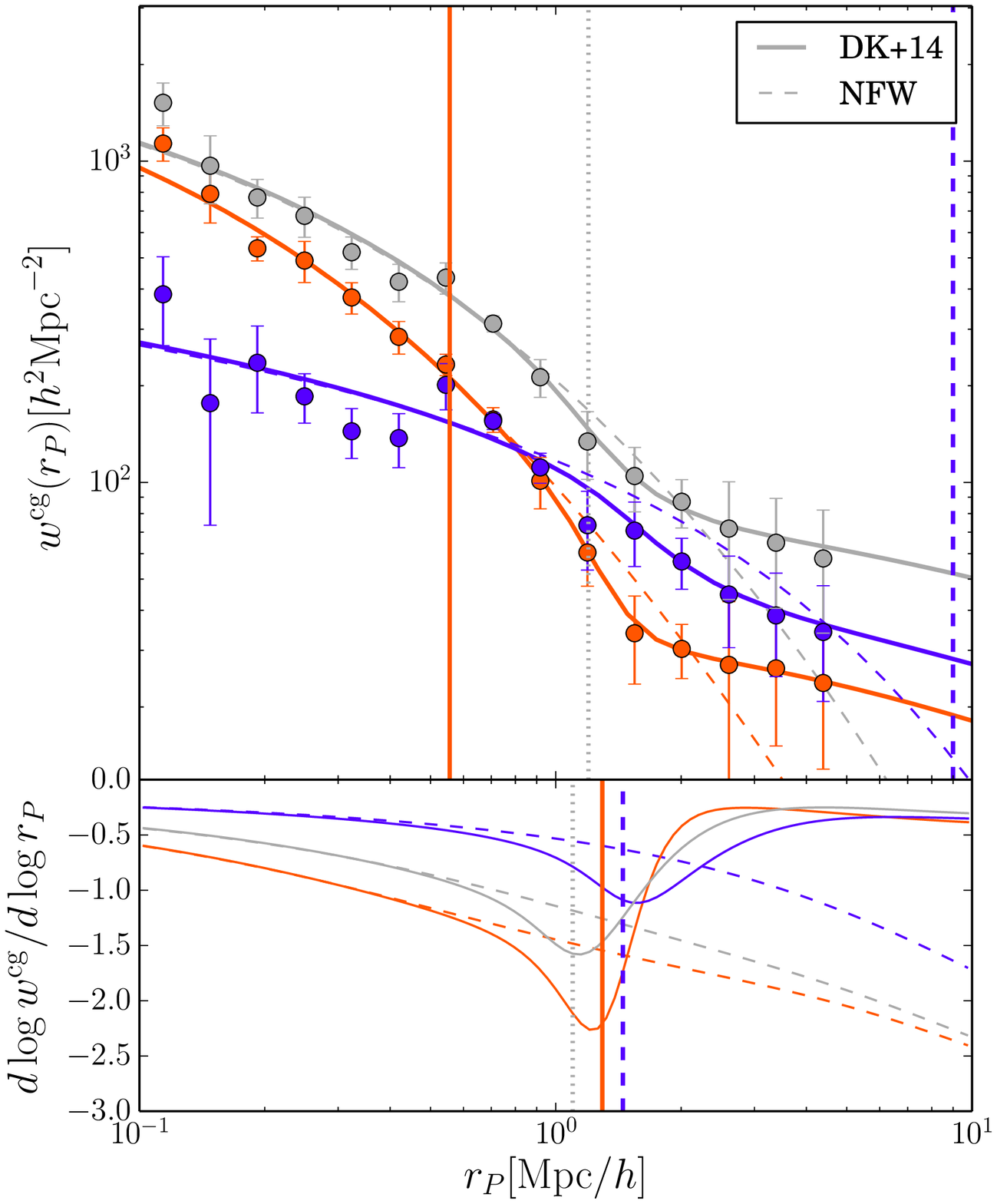}\\
    \includegraphics[width=0.5\linewidth]{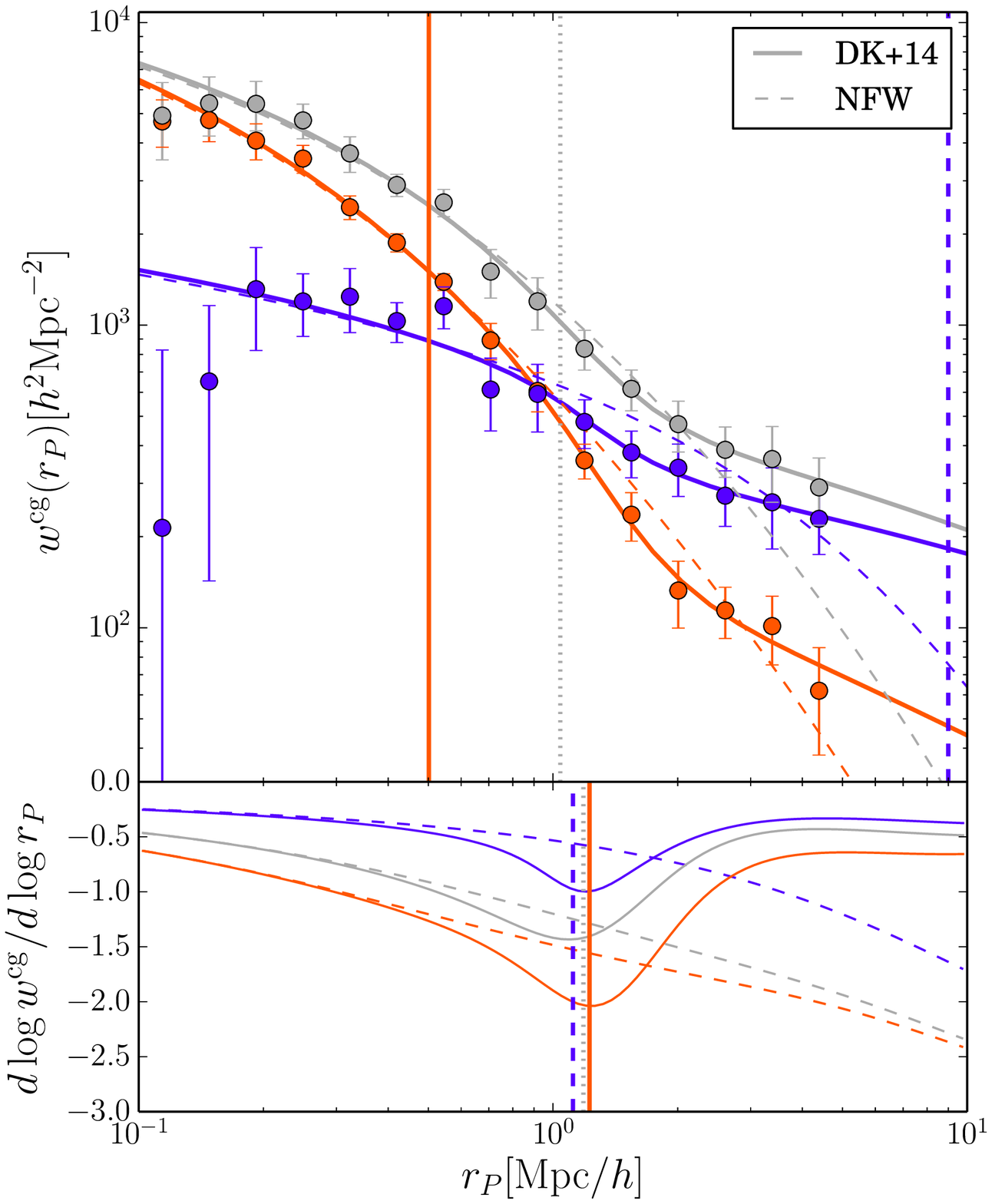}&
    \includegraphics[width=0.5\linewidth]{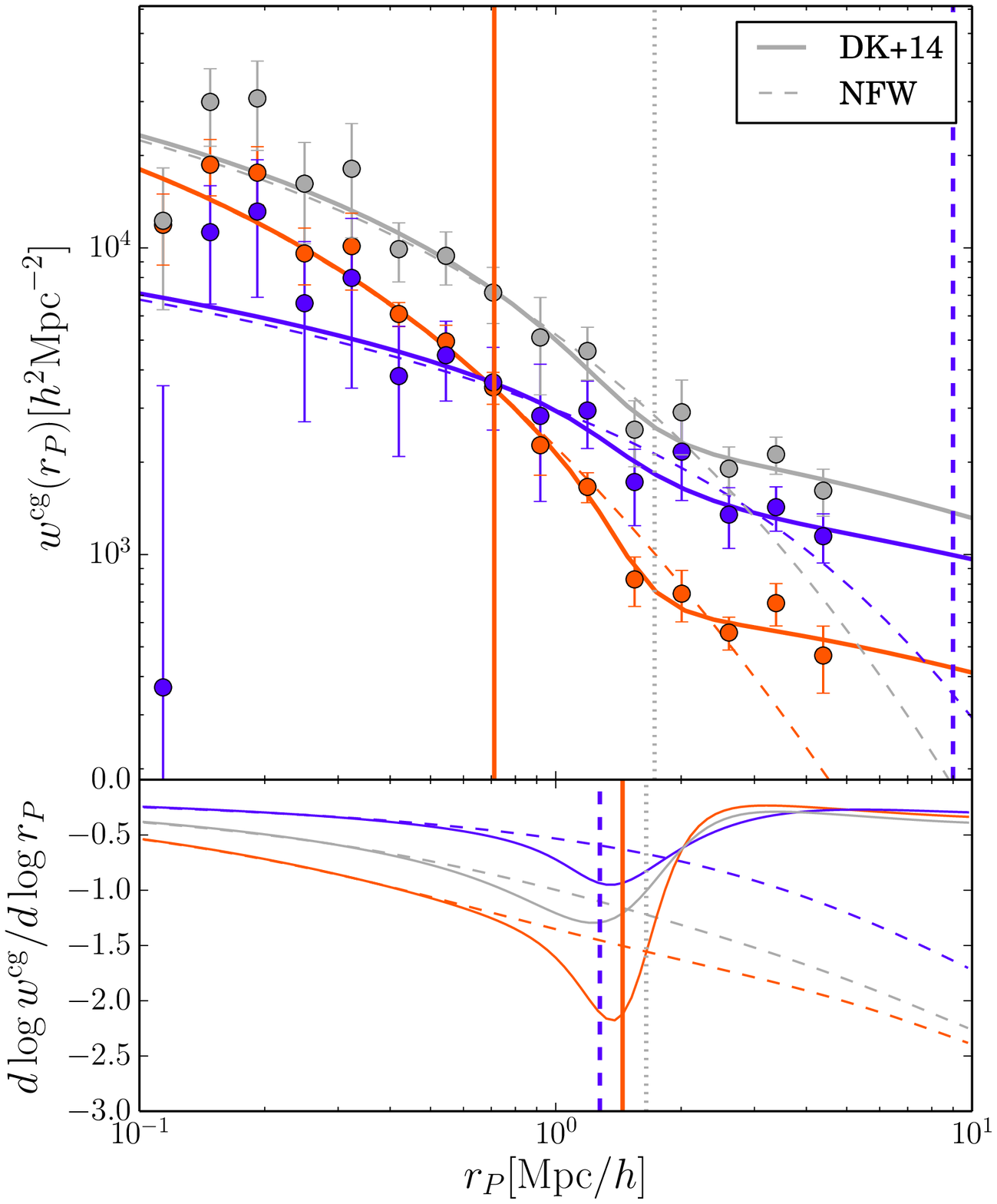}\\
\end{tabular}
  \caption{Radial number density profiles of galaxies around clusters.
    From left to right, the mean redshifts of clusters are 0.18, 0.33,
    0.55, and 0.85. Red and blue symbols show profiles for red and
    blue member galaxies. Gray symbols show profiles for all
    galaxies. Dashed lines are best-fit models of the projected NFW
    profile.  Solid lines show the best-fit \citet{DK:2014}
    model. Vertical solid and dashed lines indicate the best-fit scale
    radii $r_s$ for red and blue galaxies, respectively. Bottom panels
    show the slope of the best-fit profiles with vertical lines being
    the best-fit value of $r_t$, which can be compared with the 
    splashback radius, $R_{\rm sp}$.
    \label{fig:profiles}}
\end{figure*}

Figure~\ref{fig:profiles} show the radial number density profiles for
red and blue components at redshifts $z=0.18$, $0.33$, $0.55$, and $0.85$.
The covariance matrices are estimated from the jackknife resampling as
\begin{eqnarray}
  \widehat{{\rm Cov}_{ij}}
  &=&
  \frac{N_a-1}{N_a}\sum_a 
  [w_a^{\rm cg}(r_{{\rm p}, i}) - \overline{w^{\rm cg}}(r_{{\rm p}, i})]\times
  \nonumber \\
  &&\hspace{6em}
  [w_a^{\rm cg}(r_{{\rm p}, j}) - \overline{w^{\rm cg}} (r_{{\rm p}, j})],
\end{eqnarray}
where $\overline{w^{\rm cg}}$ is the arithmetic mean of $w^{\rm cg}_a$.  We
divide the entire area into 35 rectangular regions with a side of $5.0$~deg.
That scale corresponds to the comoving angular separation of $25 \Mpch$ at
$z=0.1$ which is sufficiently larger than the scale of our interest and
includes more than one cluster in all sub-divided regions.  We first
fit to an NFW profile by restricting to scales below $1 \Mpch$.  Best-fit scale
radii are summarized in the Table~\ref{tab:fit_prof1}. We find notable differences
of concentrations between two populations. Red galaxies are more concentrated
toward the cluster center (i.e., smaller $r_s$) and blue galaxies are less
concentrated (i.e., larger $r_s$). The difference in the concentration can be
accounted for by the merger of clusters as discussed in
\cite[e.g][]{OkamotoNagashima:2003, OkamotoNagashima:2001}.  Another possible
explanation is that red galaxies at the same luminosity live in more massive
halos than their blue counterparts \citep{Mandelbaum:2006,
More:2011} and have experienced more dynamical friction so that they are
concentrated toward the cluster center as we see in the discussion below.  If
we focus on the redshift evolution, it is seen that the overall profile tends
to be more concentrated for lower redshifts.  While the evolution of the
concentration of red galaxies is subtle, blue galaxies evolve rapidly from
$z=0.5$ to $0.3$.

Next we fit all the data to the full profile of equation~(\ref{eq:fullprof1}),
keeping $r_s$ and $\rho_s$ fixed to their best fit values from the simple NFW
profile fitting to simplify the degeneracies inherent in the fitting procedure.
The best fit curves are presented with solid lines in
Figure~\ref{fig:profiles}. We also show the logarithmic slope of the profile,
which can be used to define the splashback radius $R_{\rm sp}$ as a local minimum
of the slope.  We compare the splashback radii $R_{\rm sp}$, $r_t$, and $r_s$
in Table~\ref{tab:fit_prof1}.

\begin{table*}[th]
  \begin{center}
    \begin{tabular}{cc|ccccc}\hline\hline
      $z$    &sample & $r_s$ &$r_t$   & $R_{\rm sp}$ & $\Delta$AIC & $\Delta$BIC\\ \hline
                &red   & 0.50$\pm$0.01 
                           & 1.20$\pm$0.13
                           & 1.21$\pm$0.70
                                                     & -15.1 & -10.9 \\
      0.18   &blue & 1.14$\pm$0.03  
                           & 3.63$\pm$1.07 
                           & 2.66$\pm$1.29
                                                     &  -16.3 & -12.2 \\
                &all    & 0.57$\pm$0.01   
                          & 1.11$\pm$0.13 
                           & 1.21$\pm$0.66
                                                     & -15.3 & -11.1 \\ \hline
                &red   & 0.51$\pm$0.01  
                           &1.34$\pm$0.16 
                           & 1.26$\pm$0.66
                                                     & -16.2 & -12.1 \\
     0.33    &blue & 8.61$<$  
                          & 1.43$\pm$0.18 
                           & 1.52$\pm$0.58
                                                     & -17.1 & -12.9 \\    
                &all    &0.91$\pm$0.02  
                          & 1.53$\pm$0.22 
                           &1.15$\pm$0.64
                                                     & -16.6 & -12.4 \\ \hline
                &red   & 0.54$\pm$0.01 
                           & 1.20$\pm$0.10 
                           & 1.26$\pm$0.73
                                                     & -15.7 & -11.6 \\
     0.55    &blue &8.42 $<$ 
                          &  1.20$\pm$0.18 
                          & 1.26$\pm$0.54
                                                     &  -16.1 & -11.9 \\    
                &all    & 1.11$\pm$0.03   
                           &1.12$\pm$0.12 
                           & 1.10$\pm$0.52
                                                     & -15.5 & -11.3 \\ \hline
                &red   & 0.77$\pm$0.03 
                           &1.40$\pm$0.13 
                           & 1.32$\pm$0.79
                                                     & -17.2 & -13.1 \\ 
      0.85   &blue & 7.73 $<$ 
                           &1.75$\pm$0.47 
                           & 1.67$\pm$0.54
                                                     & -17.3 & -13.1 \\    
                &all    &1.72$\pm$0.12
                           & 1.29$\pm$0.27 
                           &1.21$\pm$0.39
                                                     & -17.3 & -13.2 \\ \hline\hline
    \end{tabular}
  \end{center}
  \caption{Best-fit values of cluster profile parameters.
    For each redshift range, the top, middle and bottom rows are for 
    red galaxy, blue galaxy and all galaxy samples, respectively.
    All the values are in comoving $\Mpch$.
    Also shown are difference of information criteria. Minus value
    means that the NFW profile is favored over density jump model.
    \label{tab:fit_prof1}  }
\end{table*}

As shown in \citet[][]{More+:2015},
there is a tight relation between the mass accretion rate of clusters and the
normalized splashback radius $R_{\rm sp}/R_{200m}$ obtained by stacking
clusters at each redshift. Given the fact that the richness limit of our
cluster sample approximately corresponds to a constant mass limit of
$M_{200m}>10^{14} \Msun$ over the whole redshift range \citep{Oguri+:2017}, we
find that the splashback radii from our fits roughly
correspond to $R_{\rm sp} \sim R_{\rm 200m}$.
This is broadly consistent with \citet{More+:2016} in which
splashback radii were derived for Sloan Digital Sky Survey (SDSS)
clusters to argue that the observed splashback radii are smaller than the
standard cold dark matter model prediction, implying the faster mass accretion
than the standard model. However more careful estimates of the mass of these
clusters using the weak gravitational lensing signal ought to be performed
before doing a more quantitative comparison to \citet{More+:2016} as well as to
quantify the redshift evolution. We will explore this in the near future.


We compare the goodness of fit between NFW and full profile of
\cite{DK:2014} by computing two different criteria; Akaike Information
Criteria (AIC) corrected for the finite data size and Bayesian Information
Criteria (BIC). They are defined as
\begin{eqnarray}
  \displaystyle
  {\rm AIC} &=& -2\ln({\mathcal{L}}) + 2p + \frac{2p(p+1)}{N-p-1} \\
  {\rm BIC} &=& -2\ln({\mathcal{L}}) + \ln(N)p,
\end{eqnarray}
where ${\mathcal{L}}$ is the likelihood and $p$ and $N$ are number of
parameters and data, respectively; $(N,p)$ for NFW is $(9,2)$ and 
$(15,6)$ for density jump model.
We compare the information criteria for two
profile fittings, and find that the density jump model is disfavored
as summarized in Table~\ref{tab:fit_prof1}.

As a cautionary note, we also mention that selection effects in optical cluster
finding can significantly complicate the inference of the splashback radius
from observations. Optical clusters are more likely to have their major axis
oriented along the line-of-sight which break the spherical symmetry assumption
involved in the inference of the splashback radius \citep[see
e.g.,][]{Busch:2017}. In addition degeneracies related to
cluster miscentering can also reduce the significance of the detection of the
splashback radius \citep{Baxter:2017}. The values of the splashback radii
inferred from optical clusters should therefore be carefully compared to
expectations, a topic we will focus on in the near future.

For clusters at $z>0.5$, we find a slight decline of radial profiles in
the central region, $r<0.2\Mpch$. This might reflect the mis-centering of
the optically-selected clusters, which have been inferred in
comparison with the X-ray profile \citep{Oguri+:2017}. It may be more
difficult to correctly identify the center of cluster at higher
redshift simply because in crowded regions, like cluster centers,
angular separations of neighboring galaxies are smaller at higher
redshifts given the same physical scale. This is mainly due to the
fact that the pipeline fails to deblend galaxies in 
crowded regions. We leave the effect of mis-centering for future work,
after the weak lensing measurement of the CAMIRA cluster sample becomes
available. 

\subsection{Red fraction as a function of redshift}
\label{ssec:fred-z}
\begin{figure}[th]
  \begin{center}
    \includegraphics[width=\linewidth]{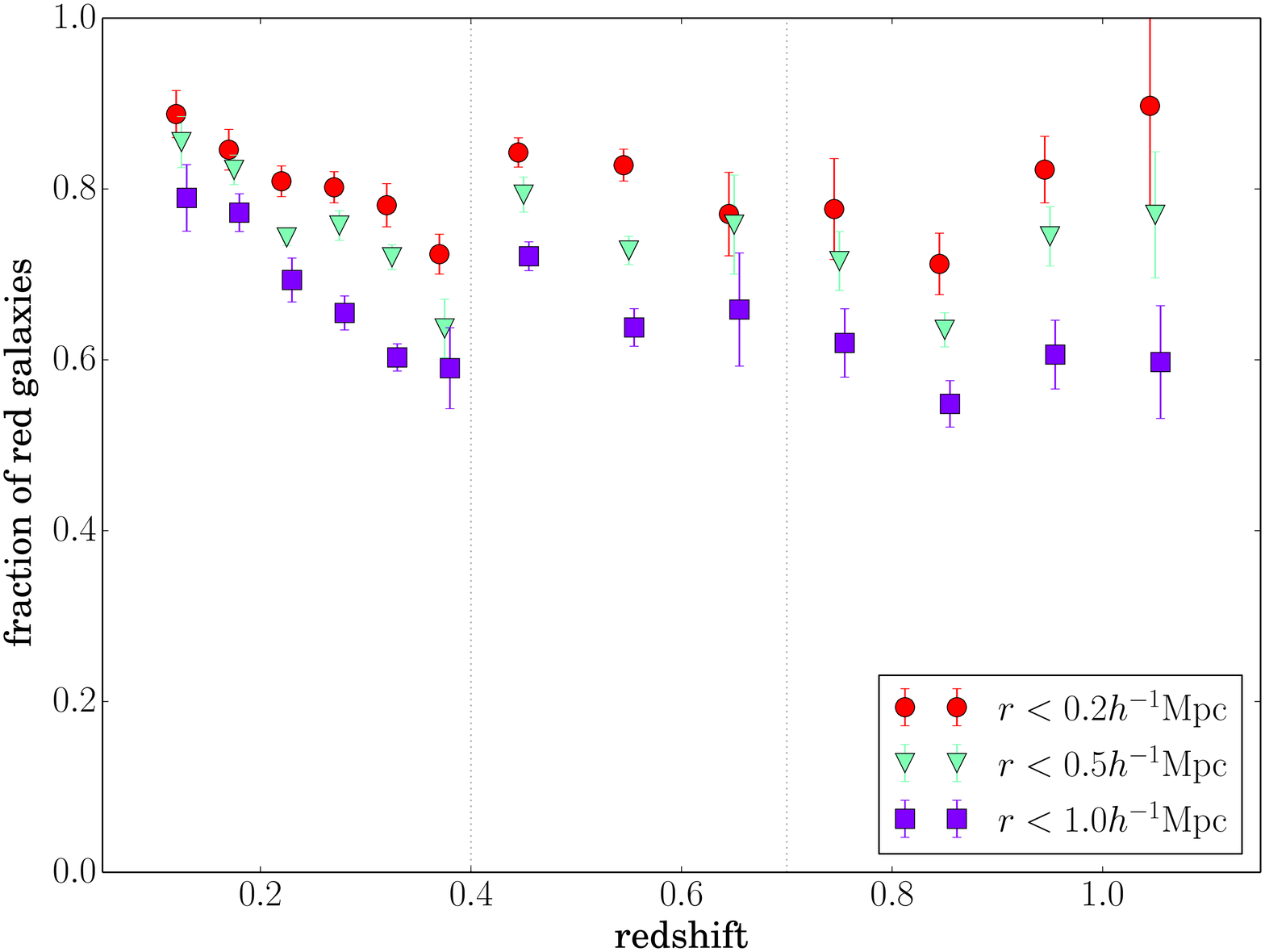}
  \end{center}
  \caption{Fraction of the red galaxies as a function of redshift and
    the maximum distance from the cluster center for counting up the
    galaxies. The x-axis for different $r_P$ is slightly shifted for
    visual purpose.
    \label{fig:fred_rz}}
\end{figure}

We derive the red galaxy
fraction by summing up the number of red and blue galaxies in each
radial bin out to the maximum comoving distances, for which we adopt
$0.2$, $0.5$, and $1.0\Mpch$. Figure~\ref{fig:fred_rz} shows the
evolution of fractions of red galaxies as a function of redshift, for three
different maximum distances. We can clearly see the evolution of the
red fraction over the cosmological time scale, from $z=1.1$ to $0.1$,
such that the red fraction increases at lower redshift. This
qualitative trend is consistent with \citet{Hennig+:2017,Jian+:2017},
while \citet{Loh+:2008} reported steeper evolution. The red fraction
significantly increases with decreasing maximum distance from the
cluster center, which is due to the different radial number density
profiles between red and blue galaxies.
We note that there are gaps in the fraction of red galaxies at $z=0.4$
and $z=0.75$ because we have used the different combination of filters
to define red and blue galaxies. This implies that the global evolution
of red fraction over the redshifts $0.1<z<1.1$ is subject to the choice
of colors; however, within the redshift range in which we use the same
filter combination, we observe apparent decreases of the red fraction.
However, we note that the red fraction at higher redshift bins,
$0.7<z$, are almost flat or slight increasing. As we will discuss it
at the end of the Section~\ref{ssec:simulation}, this trend is partly
due to our bad photometry at crowded region like cluster center.

\begin{figure}[th]
  \begin{center}
  \includegraphics[width=\linewidth]{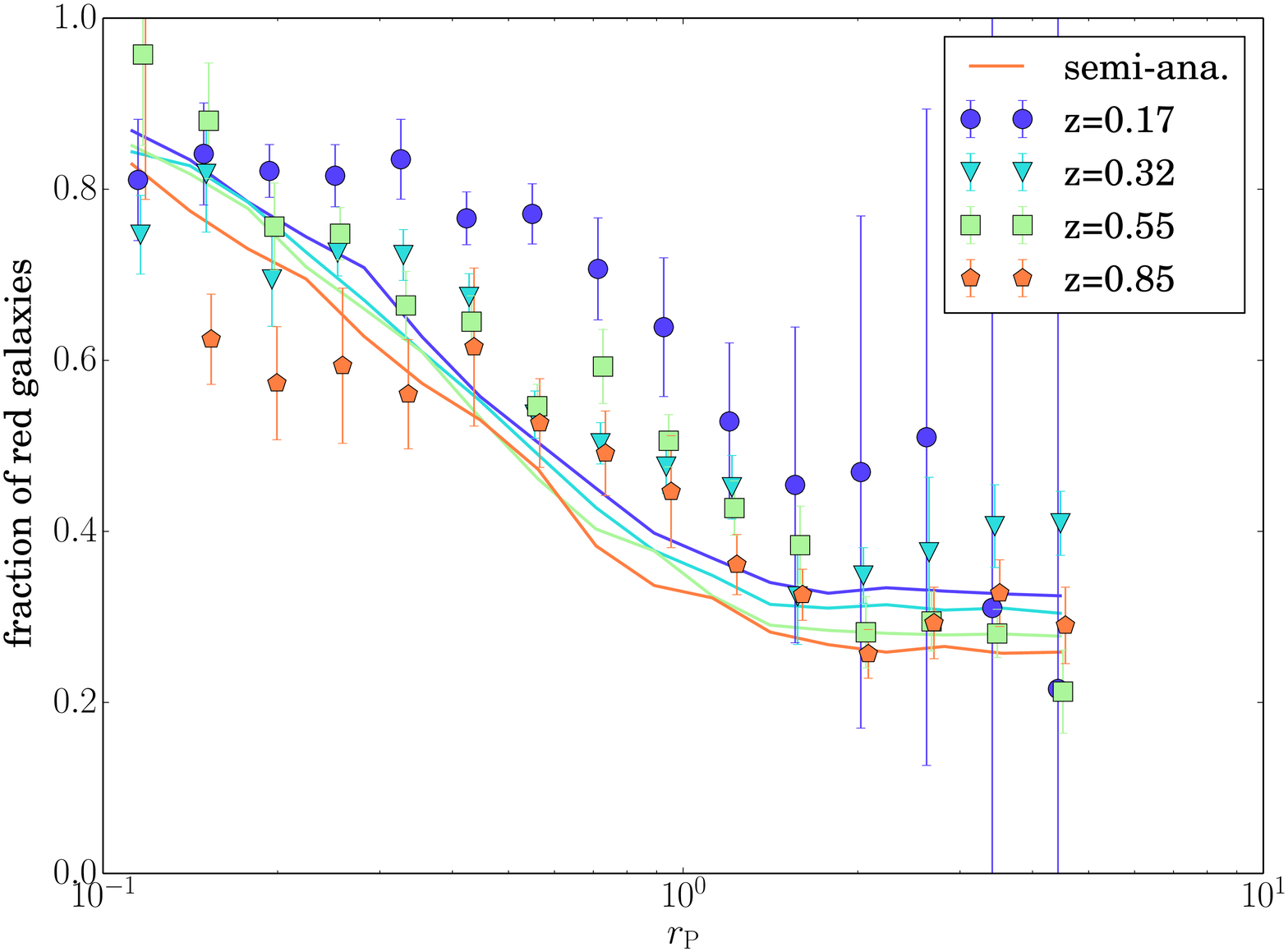}
  \caption{Radial profile of the fraction of red galaxies as a function of
    projected cluster-centric radius at four different redshifts.
    Symbols are observed data points and solid lines are the prediction
    from the semi-analytical model described in
    Section~\ref{ssec:simulation}.
  \label{fig:sim_redfraction_radial}}
\end{center}
\end{figure}

Figure~\ref{fig:sim_redfraction_radial} shows the  fraction of red
galaxies as a function of projected cluster-centric radius 
$r_{\rm P}$.
Symbols are observed points and solid lines are predictions from the semi
analytical model described in Section~\ref{ssec:simulation}.
The fractions of red galaxies are high, $\sim 0.6-1.0$ in the inner regions
($r_{\rm P}<0.3 \Mpch$), decreasing to $\sim 0.2-0.4$ in the outer
regions ($r_{\rm P}>1\Mpch$). On the intermediate scale in between the
inner and outer regions, the fraction of red galaxies is gradually
decreasing. Although the absolute value of the observed fraction is
slightly higher than predicted by the semi analytical model, the declining
slopes are in good agreement with the model.

%

%
\subsection{Comparison with Semi Analytical Model}
\label{ssec:simulation}
%
It is important to check the consistency of our results with
theoretical models of galaxy formation.  We compare our results with a
semi-analytical model of galaxy formation, $\nu^2$GC
\citep{2016PASJ...68...25M}. Semi-analytical models have an advantage
over mock galaxy catalogs based on the halo occupation distribution
technique in that semi-analytical models are constructed
from physically motivated prescriptions of several astrophysical
processes which, in comparison with observations, will lead to better
understanding of the build-up of cluster galaxies. They also contain
physical properties of galaxies such as galaxy stellar and gas masses
and star formation rates.  The information may also help reveal
physical processes affecting evolution of cluster galaxies.

We examine the spatial distribution of galaxies in $\nu^2$GC.
In
$\nu^2$GC, we construct merger trees of dark matter halos using
cosmological $N$-body simulations \citep{2015PASJ...67...61I} with the
Planck cosmology \citep{2014A&A...571A..16P}.  The simulation box is
280 $\Mpch$ on a side containing $2048^3$ particles, corresponding to
a particle mass of $2.2\times 10^8 h^{-1}M_{\odot}$. The semi-analytical
model includes the main physical processes involved in galaxy
formation: formation and evolution of dark matter halos; radiative gas
cooling and disc formation in dark matter halos; star formation,
supernova feedback and chemical enrichment; galaxy mergers; and feedback
from active galactic nuclei.  The model is tuned to fit the luminosity
functions of local galaxies \citep{Driver+:2012} and the mass function
of neutral hydrogen \citep{Martin+:2010}. The model well reproduces
observational local scaling relations such as the Tully-Fisher
relation and the size-magnitude relation of spiral galaxies
\citep{Courteau+:2007}.

We use $\nu^2$GC results to create mock galaxy catalogs. Rest-frame and
apparent magnitudes of galaxies are estimated in the same filter as 
used in the HSC survey \citep{Kawanomoto+:2017}. We apply  
the same magnitude cut of the HSC ($M_{z} <-18.5$ in rest frame) to
the simulated galaxies.  In the analysis in this paper, we extract
galaxy samples at different redshifts, residing in dark matter
halos with mass more than $10^{14} M_{\odot}$. We regard those
galaxies as cluster galaxies. 
We have $\sim 175$ simulated clusters at $z=1.1$ and $\sim 1000$
clusters at $z=0.13$ which are defined in a same realization.

Since the mock catalogs do not perfectly reproduce the colors of galaxies,
we cannot apply the selection condition that is applied to the
HSC data to the mock catalogs. We therefore re-define the selection
criterion for red and blue galaxies in the mock galaxy catalogs.
To define red and blue galaxy populations, we determine the
red-sequence using a subsample of the mock galaxies as follows. 
We perform linear regression to fit the slope $\alpha$ and zero-point
$\beta$ of the following equation to describe the red sequence
\begin{equation}
C = \alpha \ m_z + \beta,
\end{equation}
where $C$ is the color corresponding to the cluster redshift, e.g. $g-r$
at redshift [0.1--0.2].  To reduce contaminations by blue galaxies,
we only fit to galaxies with specific star formation rate (sSFR) 
$\log_{10} (\mathrm{sSFR}/\mathrm{Gyr}^{-1}) \leq -1$ 
and with distances from the cluster center $d \leq 0.1~\mathrm{Mpc/h}$.
These conditions are reasonable to extract red-sequence galaxies which
correspond to the CAMIRA-identified red sequence.  Finally, in
exactly the same manner as in the observation, we define galaxies redder than
$C - 2\sigma$ line on the CMD as red galaxies, where $\sigma$ is the
standard deviation of the distribution of the galaxies used to the
fitting.

\begin{figure*}
  \begin{tabular}{cc}
    \includegraphics[width=0.5\linewidth]{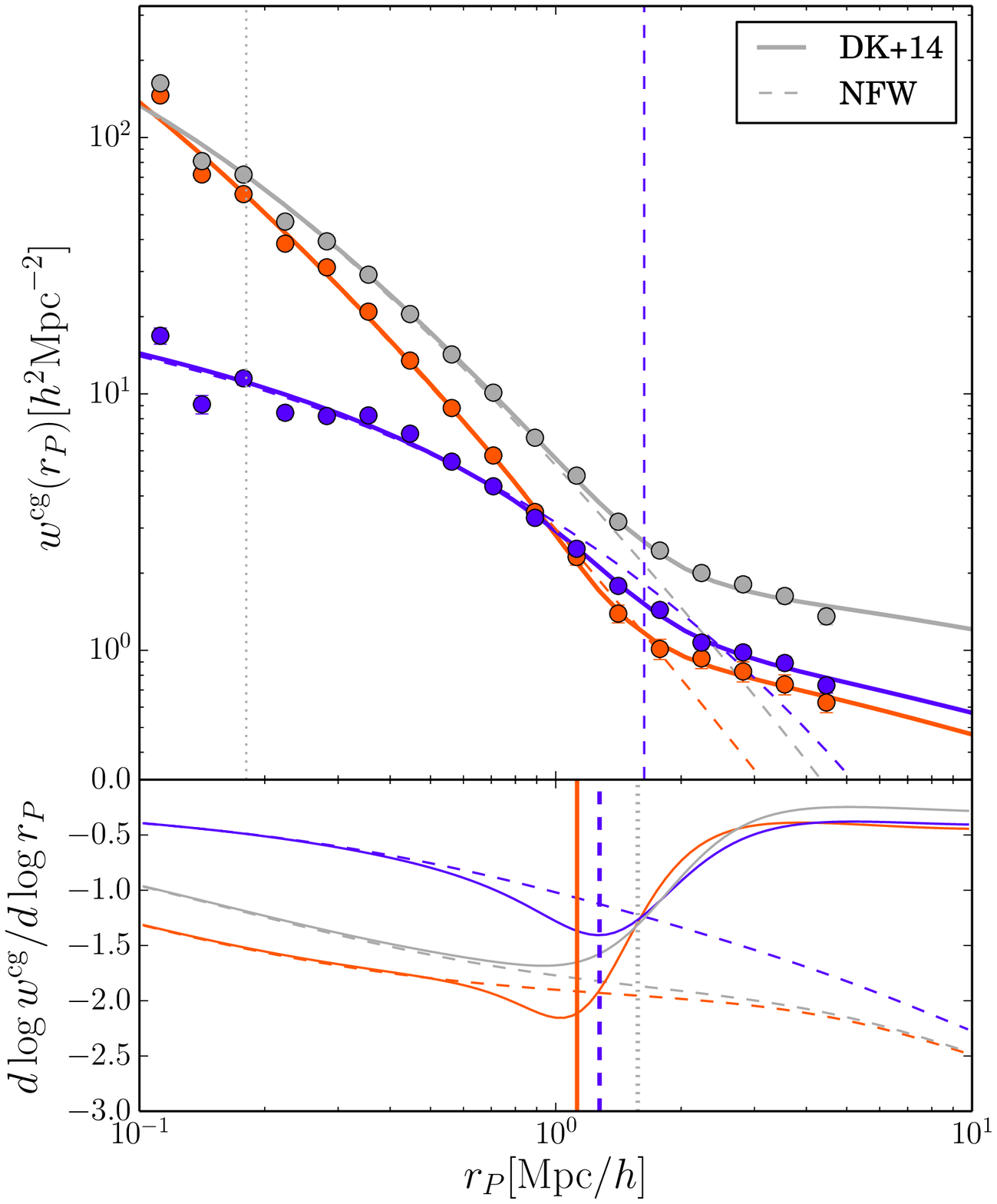}&
    \includegraphics[width=0.5\linewidth]{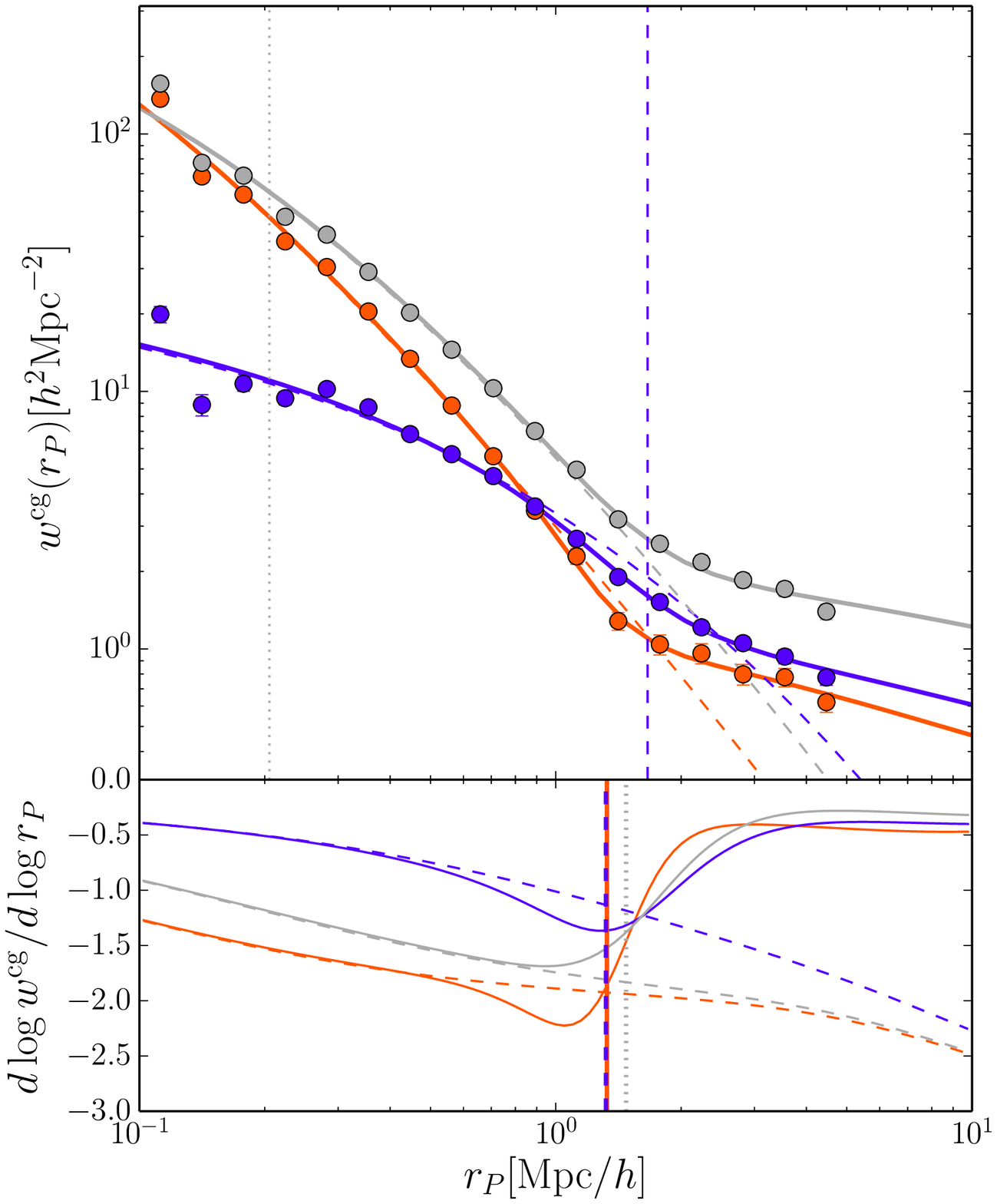}\\
    \includegraphics[width=0.5\linewidth]{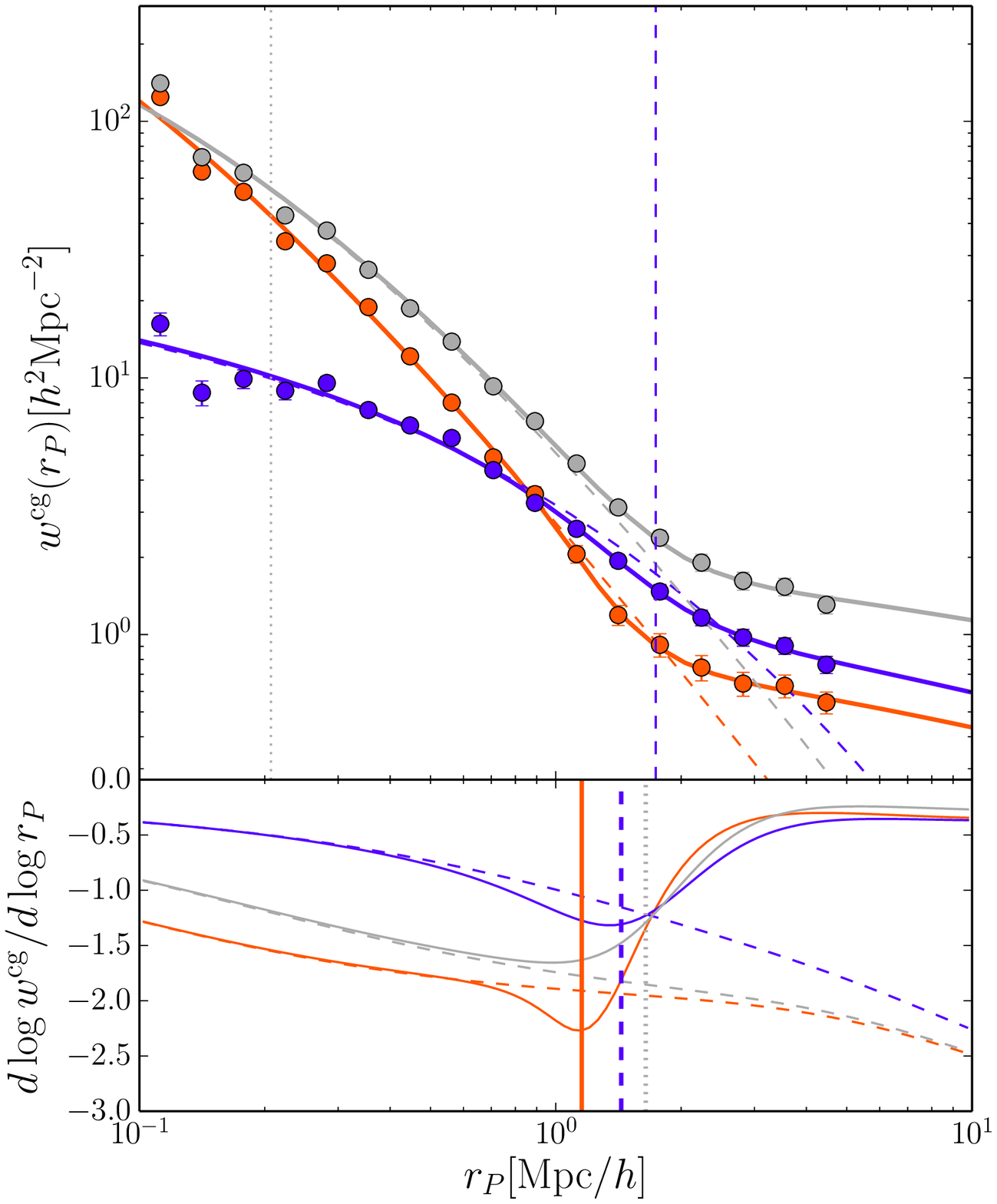}&
    \includegraphics[width=0.5\linewidth]{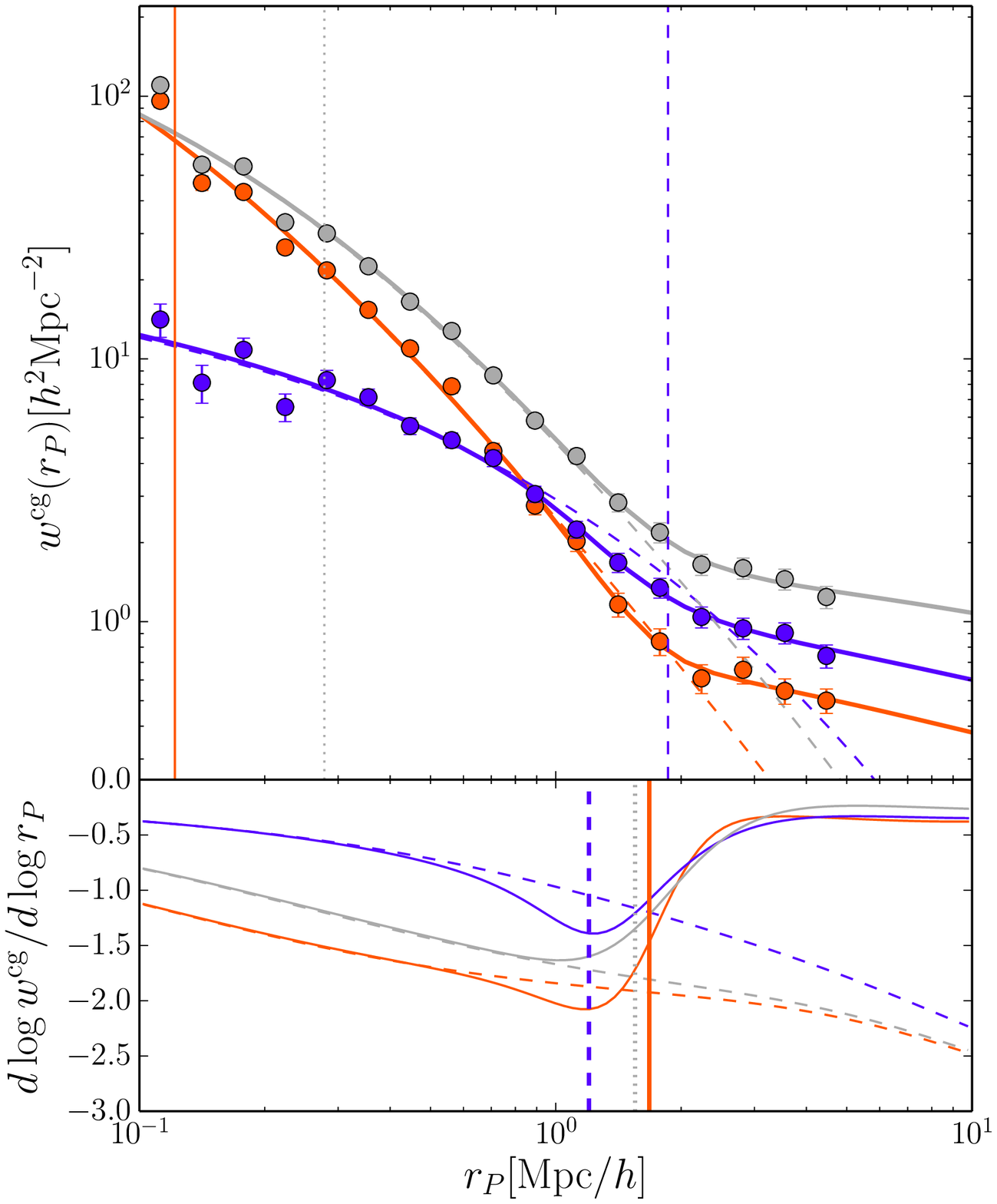}\\
  \end{tabular}
  \caption{Same as Figure~\ref{fig:profiles} but obtained from
  simulations with a semi-analytical model.
  \label{fig:profiles_sim}}
\end{figure*}
Figure~\ref{fig:profiles_sim} shows radial profiles of cluster member
galaxies for different redshift bins and different populations
identified in the simulation.
We clearly see that blue galaxies are more diffuse and
red galaxies are more concentrated, which is consistent with our
observational results. 

%

%
\begin{figure}
  \includegraphics[width=\linewidth]{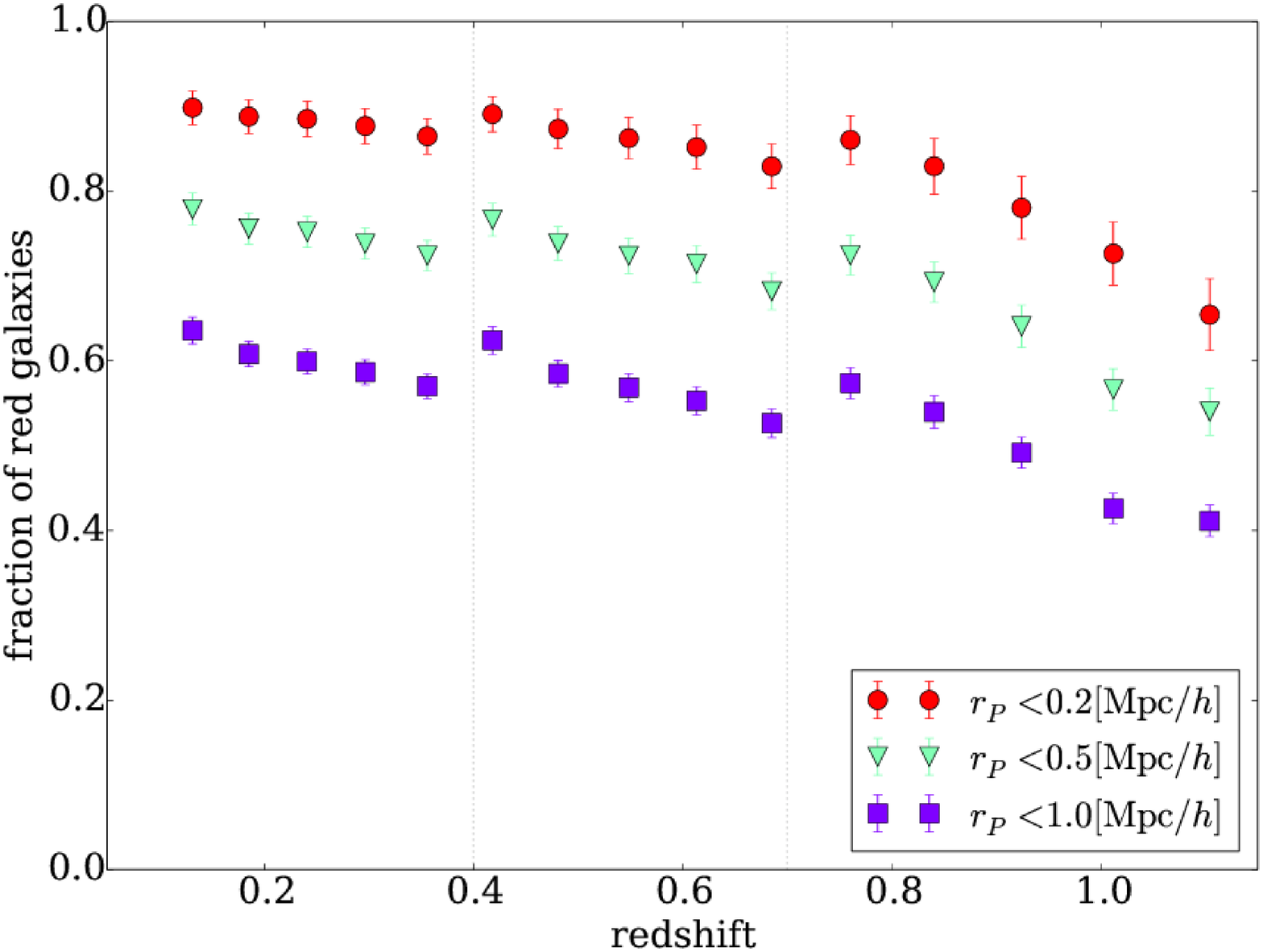}
  \caption{Same as Figure~\ref{fig:fred_rz} but obtained from
    simulations with a semi-analytical model.
  \label{fig:sim_redfraction}}
\end{figure}
Figure~\ref{fig:sim_redfraction} shows the redshift evolution of the
red fraction in the simulation with same binning of distance from the
cluster center. We see a clear decrease of the red fraction with
redshift and the result shows a reasonable agreement with the
observation except for the two highest redshift bins. 
The simulated results show a monotonic decrease of the red fraction
at the highest redshift ranges, but our observational results show a
slight increase. The monotonic decrease of the red fraction  
to $0.5$ at $z=1$ is consistent with the results of
\cite{Hennig+:2017}, and therefore the slight increase seen in our
data may not be accounted for by the difference in the depth of the
sample, (our sample is $\sim 1$ mag deeper in z-band) or 
the different filter combination used to define the red and blue
populations. We do not make strong conclusions about the source of
this discrepancy in this paper but it may be partly due to the bad
photometry in crowded regions, which significantly affects the
color of galaxies in clusters. We will revisit the issue
and address it in the future work once the photometry of HSC in the
crowded regions is improved.

%
\section{Summary}
\label{sec:summary}
%
In this paper, we have used the HSC S16A internal data release galaxy
sample over $\sim 230$~deg$^2$ to explore the properties of cluster
galaxies over the wide redshift and magnitude ranges. Clusters are
identified by the red-sequence cluster finding method CAMIRA
\citep{Oguri:2014,Oguri+:2017}. Thanks to the powerful capability of
the Subaru telescope to collect light and the good sensitivity of the
HSC detector, we can study faint cluster galaxies down to 24th mag in
$z$-band. This sample is $\sim 1$ mag deeper than the cluster
sample of \cite{Hennig+:2017} which reaches $m_\star +1.2 \sim 23.2$
in DES $z$-band at $z=1$.
Together with a reliable CAMIRA cluster catalog out to
$z=1.1$, the excellent HSC data allows us to continuously track the
evolution history of cluster galaxies from $z=1.1$ to the present. 

We have used the stacked color-magnitude diagram to divide red and
blue galaxies in clusters. We have statistically subtracted background
and foreground galaxies after area corrections using the well defined
random catalog, which are also available from the HSC database. After
subtracting the foreground and background galaxies, color-magnitude
relations for red galaxies (red-sequence) and blue clouds are clearly
detected over  wide ranges in redshift and magnitude. We have used
these color-magnitude diagrams for defining red and blue galaxies,
studying the tightness of the red-sequence down to very faint
magnitudes, and the radial number density profiles of red and blue
galaxies. Our results are summarized as follows.
\begin{itemize}
\item{} Red galaxies in clusters follow a clear linear relation in the
  color-magnitude diagram down to the HSC completeness limit for all
  redshifts. 
\item{} However, we observe a slight offset of the red populations in
  the cluster from the linear relation determined by the CAMIRA member
  galaxies.  This may be partly due to our sample selection,
  i.e. a constant mass cut over all redshift ranges and we will
  revisit it once the cluster mass is  well measured with the weak lensing.
\item{} We have measured the intrinsic scatter of the red-sequence as
  a function of the observed $z$-band magnitude and cluster redshift. We
  have found that the intrinsic scatter is almost constant over wide
  range of magnitudes. The intrinsic scatter shows little evolution
  with redshift.
\item{} Red galaxies are more concentrated toward the cluster center
  compared with blue galaxies. We have fit the cluster member
  radial profile at
  $r<1.0\Mpch$ to an NFW profile, and find the transition scale $r_s$ 
  is significantly smaller for red galaxies than for blue
  galaxies. Given that the cluster sample has approximately constant
  mean mass over different redshifts \citep{Oguri+:2017}, the mildly decreasing
  $r_s$ with redshift implies that the galaxy profiles in
  clusters become less concentrated at higher redshift.
  We note, however, that it is important to independently
  derive the virial mass of the clusters; this will be
  soon provided by the HSC weak lensing analysis
  \citep{Mandelbaum+:2017}. The special care is
  required that the mass profile measured by weak lensing is for dark
  matter and this should be different from the profile of member galaxies.
\item{} We have fit the radial number density profiles with the
  density jump model of \cite{DK:2014}, and found that the splashback
  radius $R_{\rm sp}$ defined by the minimum of logarithmic slope is
  almost constant over the redshift range. 
  However, given the large statistical uncertainties, we do not detect
  the splashback radius for our current data set.
\item{} The fraction of red galaxies is not only a strong function
  of the distance from the cluster center but also exhibits a moderate
  decrease with increasing redshift. We note that the estimated red
  fraction shows a slight discontinuity at the redshift where the red
  and blue galaxies are defined in different combination of colors,
  i.e. $z\sim 0.4$ and $z\sim 0.7$. This discontinuity might reflect
  that our definition of red and blue galaxies are not optimal near
  the transition redshifts because the redshift of the 4000$\AA$ break
  mismatches to the filter response functions of given combination of
  the color. We note that the same discontinuity is seen also in the
  simulations.
\item{} We have also compared our results with 
  semi-analytical model predictions. We have found that the observed
  cluster profiles and the redshift evolution of the red fraction are
  broadly consistent with the semi-analytical model prediction. Further
  studies for more quantitative comparisons are important.
\end{itemize}

The total mass and mass profile of the CAMIRA clusters can be measured
by stacked weak lensing. With the help of the mass-richness relation
by the forward modeling \citep{Murata+:2017}, this will allows us to
explore more in detail of cluster physical quantities such as virial
radius, mass accretion rate and mass dependence of those quantities. 
We will revisit this in our future work.



\begin{ack}
We thank the anonymous referee to provide useful comments.
AN is supported in part by MEXT KAKENHI Grant Number 16H01096.
This work was supported in part by World Premier International
Research Center Initiative (WPI Initiative), MEXT, Japan, 
MEXT as ``Priority Issue on Post-K computer'' (Elucidation of the
Fundamental Laws and Evolution of the Universe) and JICFuS, 
and JSPS KAKENHI Grant Number 26800093 and 15H05892. 
SM is supported by the Japan Society for Promotion of
Science grants JP15K17600 and JP16H01089.
This work was supported in part by MEXT KAKENHI Grant Number 17K14273
(TN).
HM is supported by the Jet Propulsion Laboratory, California Institute
of Technology, under a contract with the National Aeronautics and
Space Administration. 

The Hyper Suprime-Cam (HSC) collaboration includes the astronomical
communities of Japan and Taiwan, and Princeton University.
The HSC instrumentation and software were developed by the National
Astronomical Observatory of Japan (NAOJ), the Kavli Institute for the
Physics and Mathematics of the Universe (Kavli IPMU), the University
of Tokyo, the High Energy Accelerator Research Organization (KEK), the
Academia Sinica Institute for Astronomy and Astrophysics in Taiwan
(ASIAA), and Princeton University.  Funding was contributed by the FIRST 
program from Japanese Cabinet Office, the Ministry of Education, Culture, 
Sports, Science and Technology (MEXT), the Japan Society for the 
Promotion of Science (JSPS),  Japan Science and Technology Agency 
(JST),  the Toray Science  Foundation, NAOJ, Kavli IPMU, KEK, ASIAA,  
and Princeton University.

The Pan-STARRS1 Surveys (PS1) have been made possible through
contributions of the Institute for Astronomy, the University of
Hawaii, the Pan-STARRS Project Office, the Max-Planck Society and its
participating institutes, the Max Planck Institute for Astronomy,
Heidelberg and the Max Planck Institute for Extraterrestrial Physics,
Garching, The Johns Hopkins University, Durham University, the
University of Edinburgh, Queen's University Belfast, the
Harvard-Smithsonian Center for Astrophysics, the Las Cumbres
Observatory Global Telescope Network Incorporated, the National
Central University of Taiwan, the Space Telescope Science Institute,
the National Aeronautics and Space Administration under Grant
No. NNX08AR22G issued through the Planetary Science Division of the
NASA Science Mission Directorate, the National Science Foundation
under Grant No. AST-1238877, the University of Maryland, and Eotvos
Lorand University (ELTE).

This paper makes use of software developed for the Large Synoptic
Survey Telescope. We thank the LSST Project for making their code
available as free software at http://dm.lsst.org.

\end{ack}


\bibliographystyle{mn2e}
\bibliography{bibdata}

\end{document}